\documentclass{emulateapj}
\newcommand{\cha}{\textit{Chandra }}
\def\xmm{{XMM-{\it Newton\/ }}}

\def\leg{{\em COSMOS-Legacy }\/}

\shorttitle{The \textit{Chandra} COSMOS Legacy survey: the \MakeLowercase{z}$>$3 sample}
\shortauthors{Marchesi et al.}

\usepackage{float}
\usepackage{color}
\usepackage{graphicx}
\usepackage{gensymb}
\usepackage{enumitem}

\begin{document}


\title{The \textit{Chandra} COSMOS Legacy survey: the $\MakeLowercase{z}>$3 sample}

\author{S. Marchesi\altaffilmark{1,2,3}, F. Civano\altaffilmark{1,3}, M. Salvato\altaffilmark{4}, F. Shankar\altaffilmark{5}, A. Comastri\altaffilmark{2}, M. Elvis\altaffilmark{3}, G. Lanzuisi\altaffilmark{2,6}, B. Trakhtenbrot\altaffilmark{7}, C. Vignali\altaffilmark{2,6}, G. Zamorani\altaffilmark{2}, V. Allevato\altaffilmark{8}, M. Brusa\altaffilmark{2,6}, F.Fiore\altaffilmark{9}, R. Gilli\altaffilmark{2}, R. Griffiths\altaffilmark{10}, G. Hasinger\altaffilmark{11}, T. Miyaji\altaffilmark{12}, K. Schawinski\altaffilmark{7}, E. Treister\altaffilmark{13,14}, C.M. Urry\altaffilmark{1}}

\altaffiltext{1}{Yale Center for Astronomy and Astrophysics, 260 Whitney Avenue, New Haven, CT 06520, USA}
\altaffiltext{2}{INAF--Osservatorio Astronomico di Bologna, via Ranzani 1, 40127 Bologna, Italy}
\altaffiltext{3}{Harvard Smithsonian Center for Astrophysics}
\altaffiltext{4}{Max-Planck-Institut f{\"u}r extraterrestrische Physik, Giessenbachstrasse 1, D-85748 Garching bei M{\"u}nchen, Germany}
\altaffiltext{5}{Department of Physics and Astronomy, University of Southampton, Highfield, SO17 1BJ, UK}
\altaffiltext{6}{Dipartimento di Fisica e Astronomia, Universit\`a di Bologna, viale Berti Pichat 6/2, 40127 Bologna, Italy}
\altaffiltext{7}{Institute for Astronomy, Department of Physics, ETH Zurich, Wolfgang-Pauli-Strasse 27, CH-8093 Zurich, Switzerland}
\altaffiltext{8}{Department of Physics, University of Helsinki, Gustaf H\"allstr\"omin katu 2a, FI-00014 Helsinki, Finland}
\altaffiltext{9}{INAF--Osservatorio Astronomico di Roma, via di Frascati 33, I-00040 Monte Porzio Catone, Italy}
\altaffiltext{10}{Physics \& Astronomy Dept., Natural Sciences Division, University of Hawaii at Hilo, 200 W. Kawili St., Hilo, HI 96720, USA}
\altaffiltext{11}{Institute for Astronomy, 2680 Woodlawn Drive, University of Hawaii, Honolulu, HI 96822, USA}
\altaffiltext{12}{Instituto de Astronom\'ia sede Ensenada, Universidad Nacional Aut\'onoma de M\'exico, Km. 103, Carret. Tijunana-Ensenada, Ensenada, BC, Mexico}
\altaffiltext{13}{Universidad de Concepci\'{o}n, Departamento de Astronom\'{\i}a, Casilla 160-C, Concepci\'{o}n, Chile}
\altaffiltext{14}{Pontificia Universidad Cat\'{o}lica de Chile, Instituto de Astrofisica, Casilla 306, Santiago 22, Chile}

\begin{abstract}
We present the largest high-redshift (3 $< z <$ 6.85) sample of X-ray-selected active galactic nuclei (AGN) on a contiguous field, using sources detected in the \textit{Chandra} COSMOS Legacy survey. The sample contains 174 sources, 87 with spectroscopic redshift, the other 87 with photometric redshift ($z_{phot}$). In this work we treat $z_{phot}$ as a probability weighted sum of contributions, adding to our sample the contribution of sources with z$_{phot}<$3 but $z_{phot}$ probability distribution $>$0 at $z>$3.  We compute the number counts in the observed 0.5-2 keV band, finding a decline in the number of sources at $z$$>$3 and constraining phenomenological models of X-ray background. We compute the AGN space density at $z>$3 in two different luminosity bins. At higher luminosities (log$L$(2--10 keV) $>$ 44.1 erg s$^{-1}$) the space density declines exponentially, dropping by a factor $\sim$20 from $z\sim$3 to $z\sim$6.
The observed decline is $\sim$80\% steeper at lower luminosities (43.55 erg s$^{-1}$ $<$ logL(2-10 keV) $<$ 44.1 erg s$^{-1}$), from $z\sim$3 to $z\sim$4.5. We study the space density evolution dividing our sample in optically classified Type 1 and Type 2 AGN. At log$L$(2--10 keV) $>$ 44.1 erg s$^{-1}$, unobscured and obscured objects may have different evolution with redshift, the obscured component being three times higher at $z\sim$5.  Finally, we compare our space density with predictions of quasar activation merger models, whose calibration is based on optically luminous AGN. These models significantly overpredict the number of expected AGN at log$L$(2--10 keV) $>$ 44.1 erg s$^{-1}$ with respect to our data.
\end{abstract}


\section{Introduction}
In the last two decades, the existence of a co-evolutionary trend between active galactic nuclei (AGN) and their host galaxies has been established. Massive galaxies exhibit a peak in star formation at $z$$\simeq$2 (e.g., Cimatti et al. 2006; Madau and Dickinson 2014), the same redshift range ($z$=2-3) where supermassive black hole (SMBH) activity peaks, as seen in the quasar luminosity function (Hasinger et al. 2005; Hasinger 2008; Silverman et al. 2008; Ueda et al. 2014; Miyaji et al. 2015). However, before the peak in SF and AGN activity (i.e., at $z>$3) the evolution of the SMBH population is not necessarily closely linked to that of the stellar content of galaxies (Trakhtenbrot et al. 2015).

For a complete analysis of the way SMBH and galaxies co-evolve before their density peak, large samples of AGN at both high redshifts and low luminosities are required. The rest frame comoving space density of quasars at z$\geq$3 can put constraints on the BH formation scenario. The slope of the space density is linked to the time-scale of accretion of SMBHs and can therefore become a tool to investigate the SMBH formation and growth scenarios. Eventually, this may distinguish between major-merger driven accretion and secular accretion. 

Several optical surveys have already computed space densities and luminosity functions for high-z AGN ($z>$3; e.g., Richards et al. 2006; Willott et al. 2010; Glikman et al. 2011; Ross et al. 2013). However, all these works tuned their relations only at high luminosities (-27.5$<M_{AB}<$-25.5) at $z>$3, thus they all have large uncertainties in their faint end values. The limitation of optical surveys is that at low optical luminosities (-24.5$\lesssim M_{AB}\lesssim$-22), the standard color-color quasar identification procedures become less reliable, because stars can be misinterpreted as quasars. As a result, low-luminosity AGN luminosity functions from optical surveys are so far in disagreement (e.g., Glikman et al. 2011; Ikeda et al. 2011; Masters et al. 2012). Moreover, optical surveys are biased against obscured sources, whose contribution becomes also more significant at low luminosities (e.g., Ueda et al. 2014). To address both of these two issues, high-z, low luminosity X-ray selected AGN samples are required. 

Phenomenological models of AGN luminosity evolution have been developed over the years on the basis of hard X-ray (2--10 keV) surveys, with a general consensus that the  ``luminosity-dependent density evolution'' (LDDE) model describes the existing data well (Ueda et al. 2003; Hasinger et al. 2005; Gilli et al. 2007; Ueda et al. 2014; Miyaji et al. 2015; Buchner et al. 2015). The LDDE model with exponential decline (LDDEexp hereafter) shows that the peak of the AGN space density is at $z\simeq$2-3 for more luminous AGN (L$_{\rm X}>$10$^{45}$ erg s$^{-1}$), followed by an exponential decline down to $z\simeq$6. The less luminous AGN (L$_{\rm X}<$10$^{45}$ erg s$^{-1}$) in LDDE show a peak shifted towards more recent times, $z\simeq$1-2, followed by a decline to the highest redshifts reached so far ($z\simeq$3). 
An alternative model, the flexible double power-law (FDPL), has been proposed by Aird et al. (2015), as an improvement with respect to the so-called ``luminosity and density evolution'' (LADE) model (Aird et al. 2010). Although based on different assumptions, the FDPL and the LDDE models show a close agreement at all redshifts. 

However, at $z>$3 both the LDDE and the FDPL models are based on extrapolations of the low redshift predictions, given the poor statistics at these redshifts, and the space density evolution of low luminosity AGN at $z>$3 is still affected by significant uncertainties. Moreover, at $z>$3 the evolution is observed to be consistent with a pure density evolution (PDE) model, with no further corrections (Vito et al. 2014).

Physically motivated quasar activation merger models have also been developed to constrain the accretion mechanism of BH growth and to disentangle between models of BH and galaxy co-evolution. Mergers have been proposed as efficient triggering mechanisms for luminous, optically selected quasars (e.g., Barnes \& Hernquist 1991, Shen et al. 2009, Menci et al. 2014). Both phenomenological and physical models remain poorly constrained at lower luminosities at $z>$3. 

In order to put better constraints on different models, and to improve our understanding of the BH growth and of AGN triggering mechanisms in the early universe, it is necessary to improve the statistics of the low luminosity AGN population at $z>$3. 
In the last decade, several X-ray surveys (in the 2-10 keV band) have been sensitive enough to investigate this redshift range. Two pioneering studies were performed in the COSMOS field, using \xmm on the whole field (Brusa et al. 2009, $N_{\rm AGN}$=40), and \cha on the central 0.9 deg$^2$ (Civano et al. 2011, $N_{\rm AGN}$=81), reaching a luminosity limit of $L_{2-10keV}$=10$^{44.2}$ erg s$^{-1}$ and $L_{2-10keV}$=10$^{43.55}$ erg s$^{-1}$ at $z$=3, respectively. Vito et al. (2013, $N_{\rm AGN}$=34) were able to reach $L_{2-10keV}\simeq$10$^{43}$erg s$^{-1}$, using the 4 Ms \cha Deep Field South (CDF-S, Xue et al. 2011) catalog; the same group (Vito et al. 2014) studied the 2-10 luminosity function in the redshift range z=[3-5], combining deep and shallow surveys ($N_{\rm AGN}$=141). Kalfountzou et al. (2014) combined the C-COSMOS sample with the one from the wide and shallow ChaMP survey (Kim et al. 2007; Green et al. 2009; Trichas et al. 2012) to have a sample of $N_{\rm AGN}$=211 at $z>$3 and $N_{\rm AGN}$=27 at $z>$4, down to a luminosity L$_{2-10keV}$=10$^{43.55}$ erg s$^{-1}$. Finally, Georgakakis et al. (2015) combined data from different surveys to obtain a sample of  340 sources at $z>$3 over about three orders of magnitude, $L_{2-10keV}\simeq[$10$^{43}$-10$^{46}$] erg s$^{-1}$. 

In this work, we present a sample of 174 AGN with $z\geq$3 from the 2.2 deg$^2$ \cha \leg survey (Civano et al. 2016; Marchesi et al. 2016). The paper is organized as follows: in Section \ref{sec:sample_prop} we describe the sample redshift distribution, and its optical properties. In Section \ref{sec:logn-logs} we analyze the 0.5-2 keV LogN-LogS, while in Section \ref{sec:space} we use the sample to investigate 2-10 keV comoving space density in two different luminosity ranges (log$L_{\rm X}$=[43.55-44.1] and log$L_{\rm X}>$44.1), and dividing the sample in unobscured and obscured sources;  we also compare our results with previous studies and with different models of population synthesis. In Section \ref{sec:merger} we compare our results on the number density of z$>$3 AGN with detailed models of quasar activation via mergers, but also discuss possible alternatives in light of our newest data. We summarize the results and report the conclusions of our work in Section \ref{sec:conclusions}.

Throughout the paper we quote AB system magnitudes and we assume a cosmology with H$_0$ = 69.6 km s$^{-1}$ Mpc$^{-1}$, $\Omega_M$ = 0.29, and $\Omega_{\Lambda}$= 0.71.

\section{Properties of the high-redshift AGN sample}\label{sec:sample_prop}
The \cha \leg survey (Elvis et al. 2009; Civano et al. 2016) covers the 2.2 deg$^2$ of the COSMOS field, with uniform 150 ks coverage of the central 1.5 deg$^2$ and coverage between 50 and 100 ks in the external part of the field. The X-ray source catalog contains 4016 point-like sources, detected with a maximum likelihood threshold value DET$\_$ML$\geq$10.8 in at least one of three bands (0.5-2, 2-7 and 0.5-7 keV). This threshold corresponds to a probability of $\simeq$2$\times$10$^{-5}$ that a source in the catalog is actually a background fluctuation (Puccetti et al. 2009). At this threshold, the flux limit of the survey is 8.9 $\times$10$^{-16}$ in the full band (0.5-10 keV), 2.2 $\times$ 10$^{-16}$ erg s$^{-1}$ cm$^{-2}$ in the soft band (0.5-2 keV) and 1.5 $\times$ 10$^{-15}$ erg s$^{-1}$ cm$^{-2}$ in the hard band (2-10 keV). 

The catalog of optical and infrared identifications of the \cha \leg survey is presented in Marchesi et al. (2016). The source redshifts are based on spectroscopy for 2151 sources ( 54\% of the sample). For the remainder of the sample we rely on accurate photometric redshifts ($z_{phot}$) computed via best spectral energy distribution (SED) fitting as in Salvato et al. (2011; see section 2.2 for details on the accuracy).

\begin{figure*}
  \begin{minipage}[b]{.5\linewidth}
    \centering
  \includegraphics[width=1.05\linewidth]{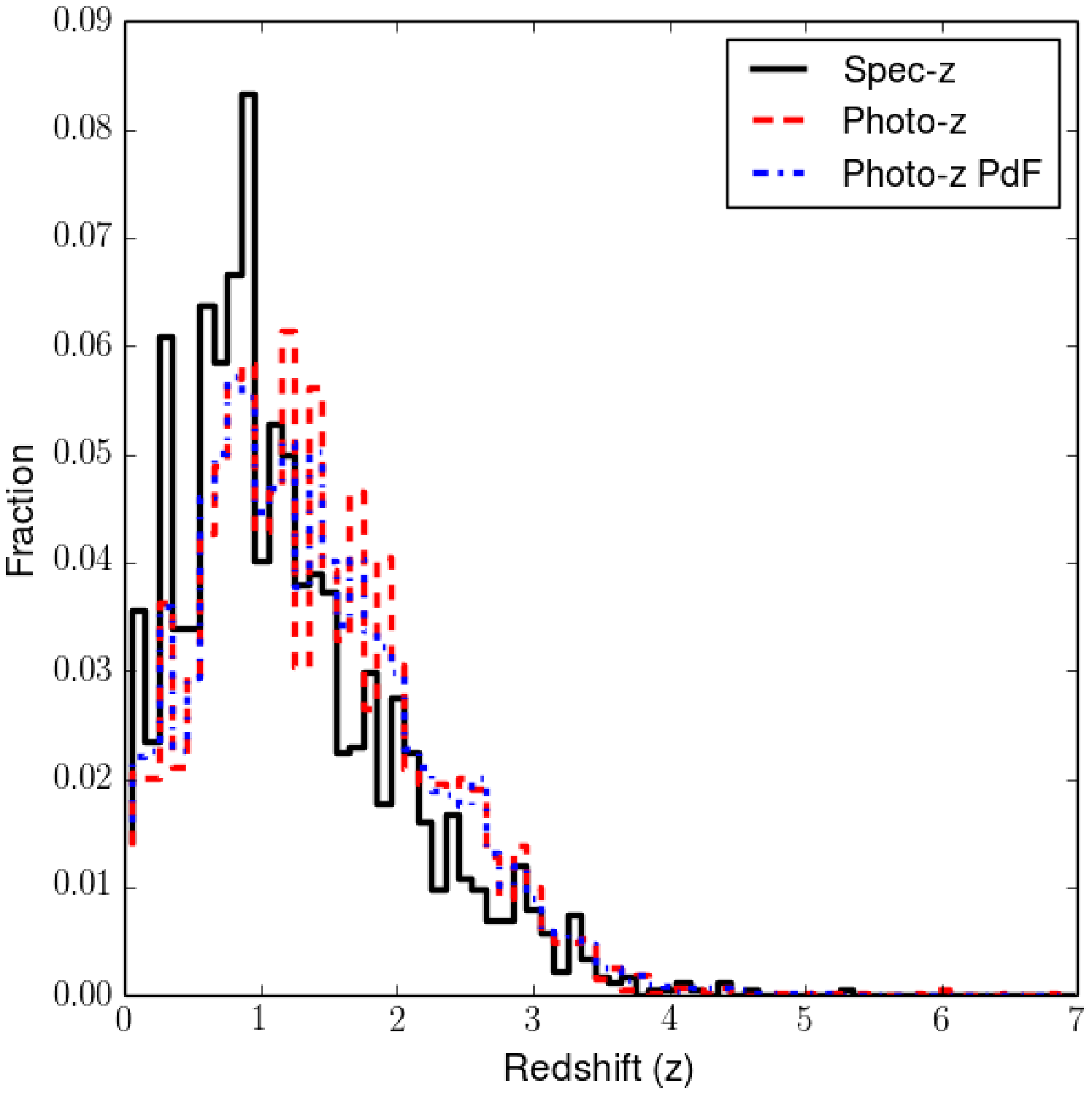}
\end{minipage}%
  \begin{minipage}[b]{.5\linewidth}
    \centering
  \includegraphics[width=0.9\linewidth]{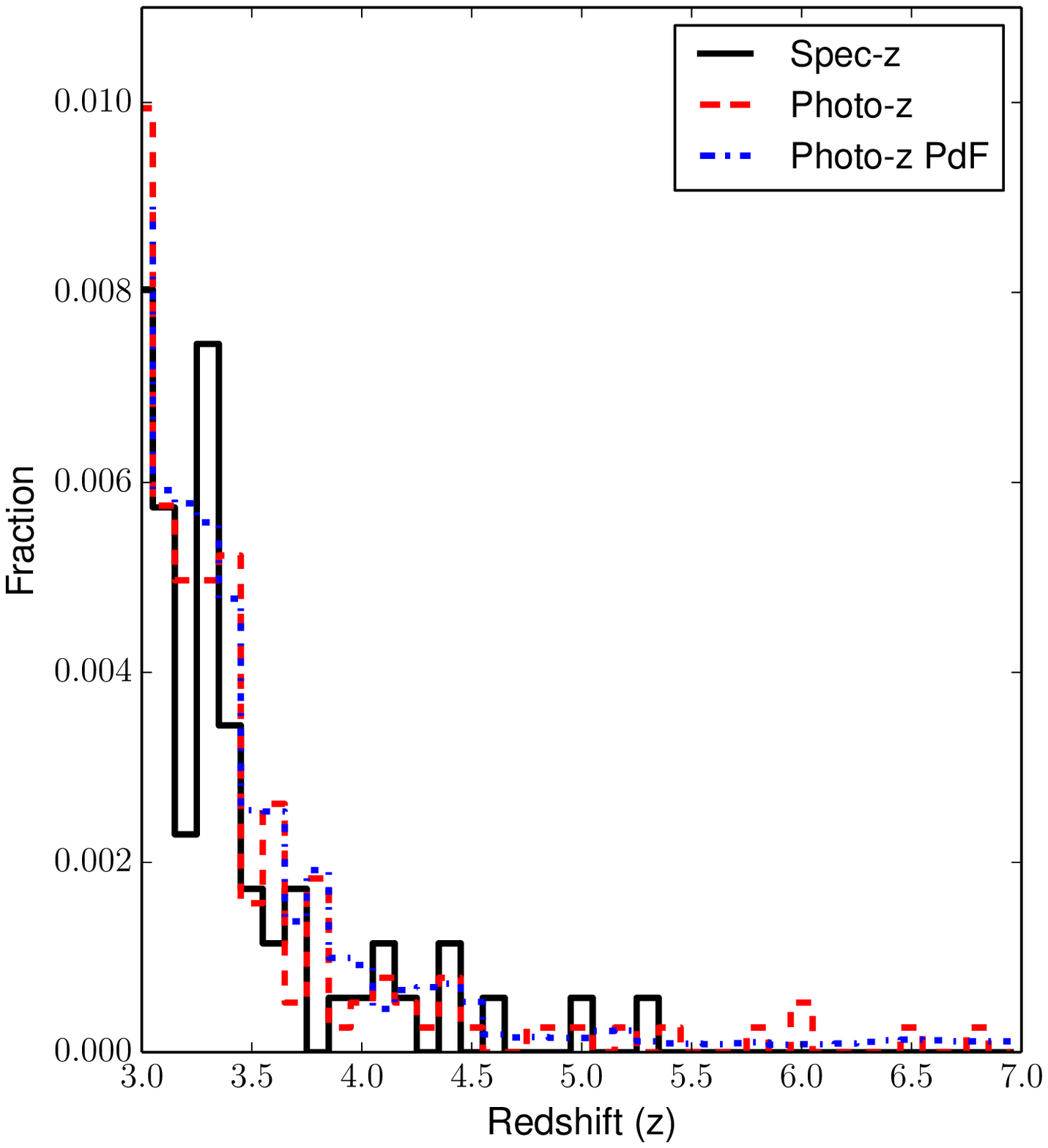}
\end{minipage}
\caption{Normalized distributions of redshift for sources with a spectroscopic redshift (black solid line), for photometric redshifts PDF peak value $z_{peak}$ (red dashed line) and for the PDF of all sources with a photometric redshift (blue dashed-dotted line), for (left) all the sources in \cha  \leg and (right) for the high redshift sample at z$\geq$3. The agreement between the photometric redshifts $z_{peak}$ distribution and the distribution of the PDFs is good at all redshifts, which means that the majority of source have narrow and highly peaked PDF.} 
\label{fig:histo_z}
\end{figure*}

\subsection{Spectroscopic redshifts}
In the \cha \leg spectroscopic sample, 87 sources have redshift greater than 3, 11 have $z\geq$4 and 2 sources have $z\geq$5. The spectroscopic redshifts were obtained with different observing programs. The zCOSMOS survey (Very Large Telescope/VIMOS; Lilly et al. 2007) and the Magellan/IMACS survey (Trump et al. 2007, 2009) are limited to $i_{AB}<$22.5. Other programs, many of which have been specifically targeting the \cha \leg sources have reached $i_{AB}$=[22.5-24.5]: these programs were carried out with \textit{Keck}-MOSFIRE (P.I. F. Civano, N. Scoville), \textit{Keck}-DEIMOS (P.I.s P. Capak, J. Kartaltepe, M. Salvato, D. Sanders, N. Scoville, G. Hasinger), \textit{Subaru}-FMOS (P.I. J. Silverman), \textit{VLT}-FORS2 (P.I. J. Coparat) and \textit{Magellan}-PRIMUS (P.I. A. Mendez).

The source with the highest spectroscopic redshift, $z$=5.3, is also the only X-ray source detected in a proto-cluster (Capak et al. 2011). Ten of the 87 sources do not have a significant detection in the soft band, and are candidate obscured objects.

\subsection{Photometric redshifts}\label{sec:photo-z}
For each \cha \leg source with optical counterpart, the photometric redshift probability distribution function (PDF) is provided\footnote{All the PDFs, together with the SED best-fit images, are available at http://irsa.ipac.caltech.edu/data/COSMOS/tables/chandra/}. PDFs are computed in steps of 0.01 up to $z$=6 and on 0.02 for 6$<$$z$$\leq$7, and to each redshift bin is associated the probability of that redshift to be the correct one. The PDF allows us to take into account sources with redshift at the PDF maximum (hereafter $z_{peak}$) $z_{peak}>$3,  but also sources with $z_{peak}<$3 that have contribution to the PDF at $z>$3. The agreement between the redshift distributions computed using  the nominal values of the photometric redshifts or the entire PDF is good at all redshifts: this can be observed in Figure \ref{fig:histo_z}, where the $z_{peak}$ histogram for photometric redshifts is plotted with a red dashed line, while the whole PDF distribution is plotted with a blue dash-dotted line. In the number counts (Section \ref{sec:logn-logs}) and space density (Section \ref{sec:space}) computation we use the PDF for each $z_{phot}$, instead that only using the $z_{peak}$ value.

The sample of sources with only photometric redshifts contains 87 sources with $z_{peak}\geq$3 (50\% of the whole sample in this redshift range), 16 sources with $z_{peak}$$\geq$4 ($\simeq$59\% of the whole sample in this redshift range), 7 sources with $z_{peak}\geq$5  ($\simeq$78\% of the whole sample in this redshift range) and 4 sources with $z_{peak}\geq$6  (100\% of the whole sample in this redshift range). The effective PDF weighted contribution at $z\geq$3 of these 87 sources is actually equivalent to having 66.0 sources with $z\geq$3, 12.8 sources with $z\geq$4, 4.7 sources with $z\geq$5 and 2.4 sources with $z\geq$6 in the sample. 

\begin{figure}
\centering
\includegraphics[width=0.5\textwidth]{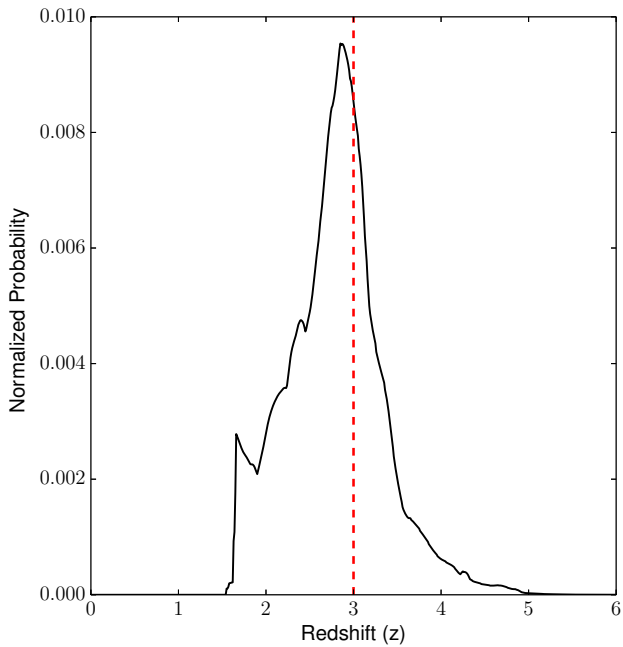}
\caption{Normalized redshift probability distribution function for source LID\_1414: this source has a $z_{phot}$ nominal value of 2.85, but has PDF$>$0 at $z$$\geq$3. The redshifts above $z$=3, weighted by their PDF, have been taken into account in the computation of the 0.5-2 keV LogN-LogS and in 2-10 keV comoving space density at $z$$>$3.}\label{fig:pdz_example}
\end{figure}

In the \cha \leg sample there are 286 sources with $z_{peak}$$<$3, but which contribute to the PDF at $z\geq$3 (i.e., with PDF$>$0 at $z_{bin}\geq$3. For example the PDF of source LID\_1414\footnote{The ``LID\_'' prefix identifies new \leg sources, while the CID\_ prefix is used for sources already in C-COSMOS}, has $z_{peak}$=2.85, but $\sim$33\% of the PDF has $z>$3 (Figure \ref{fig:pdz_example}). All these 286 sources have been taken into account in our analysis, using for each of them the contribution of each redshift bin with PDF($z_{bin}$)$>$0, weighted by the PDF value itself.

The effective contribution of these sources, i.e. the sum of all weights, is equal to add other 37.2 sources to the $z\geq$3 sample, 6.8 sources to the $z\geq$4 sample, 1.9 sources to the $z\geq$5 sample and 0.3 sources to the $z\geq$6 sample. 

In conclusion, the effective number of AGN with photometric redshifts at z$>$3 is 103.2 (66.0 with nominal $z_{phot}$$>$3 and 37.2 with nominal $z_{phot}$$<$3). A complete summary of the effective number of objects in each X-ray band and at different redshifts thresholds is shown in Table \ref{tab:sample}. Further details are provided in the sections dedicated to the analysis of the number counts (Section \ref{sec:logn-logs}) and of the space density (Section \ref{sec:space}) of our high-redshift sample.

\begin{figure}
\centering
\includegraphics[width=0.5\textwidth]{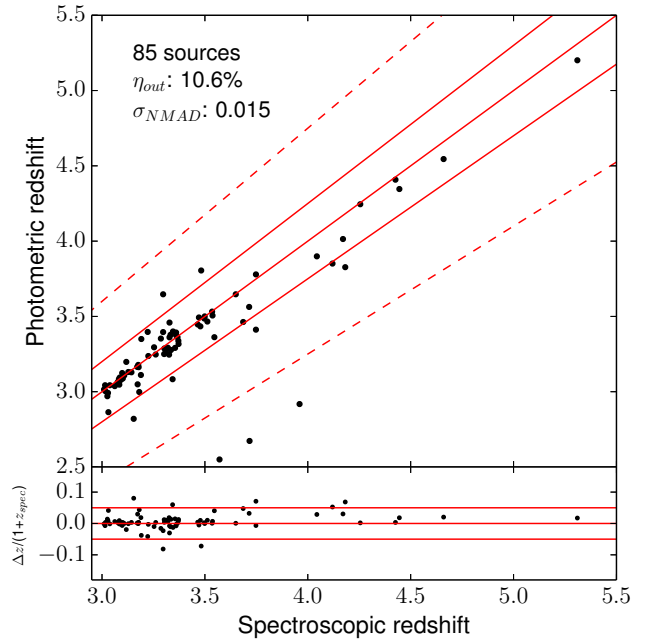}
\caption{Spectroscopic versus photometric redshift for the 85 sources in the \cha \leg sample with reliable $z_{spec}>$3 and optical magnitude information. Red solid lines correspond to $z_{phot}$ = $z_{spec}$ and $z_{phot}$ = $z_{spec}$ $\pm$0.05$\times$(1+$z_{spec}$), respectively. The dotted lines are the limits of the locus where $z_{phot}$ = $z_{spec}$ $\pm$0.15$\times$(1 + $z_{spec}$). Only three of the nine outliers, i.e., objects with $\Delta z/(1+z_{spec})>$0.15, are shown in the Figure.}\label{fig:zspec_zphot}
\end{figure}

The accuracy of the photometric redshifts established using the whole spectroscopic redshift sample is $\sigma_{\Delta z/(1+z_{spec})}$=0.02, with $\simeq$11\% of outliers $(\Delta z/(1+z_{spec})>0.15)$. At  $z$$\geq$3 there are nine outliers, but for the remaining 78 sources the agreement between $z_{spec}$ and $z_{phot}$ has the same quality of the whole sample, with a  normalized median absolute deviation ($\sigma_{NMAD}$)
\begin{equation}
\resizebox{.9\hsize}{!}{ $\sigma_{NMAD}=1.48\times median(\|z_{spec}-z_{phot}\|/(1+z_{spec}))=0.015,$} 
 \end{equation}
 (Figure \ref{fig:zspec_zphot}). 
 
As a further check, we visually inspected all the SEDs of the sources with $z_{phot}\geq$3, together with their best fits, to verify potential inaccuracies in the fit or in the SEDs data points. No source has been rejected after this visual analysis. 

Of the 87 sources with only a $z_{peak}\geq3$, 29 have no significant detection in the soft band.
The fraction of $z\geq$ AGN without a significant soft detection is significantly higher among the AGN with photometric redshift (29/87, corresponding to $\simeq$33\%) than among the AGN with spectroscopic redshift (10/87, corresponding to $\simeq$11\%). In principle, this difference in flux could be linked to different physical properties for the AGN in the two sub-samples. We will discuss this point further in section \ref{sec:opt_type}.

\subsection{Summary}\label{sec:sample_summary}
The \cha \leg sample at $z\geq$3 (L-COSMOS3) contains 174 sources with $z\geq$3, 27 with z$\geq$4, 9 with z$\geq$5, and 4 with z$\geq$6, plus other 37.2 sources with photometric redshift $z$$<$3 and contribution to the PDF at $z\geq$3. Taking into account the PDF weighted contribution, L-COSMOS3 contains 190.2 sources. 
After applying a more conservative flux cut, with a flux limit corresponding to 10\% of the \cha \leg area, the sample is reduced by 6\% and includes 179.4 sources. Such a cut in flux is applied to reduce the Eddington bias at faint fluxes (see Puccetti et al. 2009 and Civano et al. 2011 for extensive discussion on Eddington bias). This smaller sample is the one we use in the computation of the LogN-LogS (Section \ref{sec:logn-logs}) and of the space density (Section \ref{sec:space}).
This is the largest sample of X-ray selected AGN on an contiguous field, and has greater spectral completeness (50\%) than the one of other larger samples (e.g., Georgakakis et al. 2015 spectral completeness is $\sim$37\%).

A summary of the distribution of these sources in the three adopted X-ray bands versus redshift is shown in Table \ref{tab:sample}. In the same Table, we also show the number of sources used in the computation of the number counts and of the space density. 
The 0.5-2 keV observed band at $z>$3 roughly corresponds to the 2-10 keV rest-frame band: therefore in our analysis we will estimate the 2-10 keV luminosity first from the 0.5-2 keV flux, then, if the 0.5-2 keV flux is not available, from the 2-10 keV flux and, for those sources with no 2-10 keV detection, from the 0.5-10 keV flux.

At $z>$5, seven sources out of the nine in L-COSMOS3 have only a $z_{phot}$ available. The total weighted contribution of the sources, taking into account the PDF is 8.6, i.e., very close to the nominal value of nine. At $z>$5 the SED fitting is based on significantly fewer photometric points ($<$10) than at lower redshifts, and these points, mostly in the near-IR, usually have larger uncertainties than those in bluer bands. Consequently, all the results we present at $z>$5 in this work are subjected to significant uncertainties and will require a spectroscopic follow-up campaign to be confirmed or rejected.

An example of how redshift estimation of X-ray selected sources becomes complicated at high redshift can be found in three recent works focused on the search of high-z AGN in the CDF-S. Giallongo et al. (2015) adopted a NIR H band AGN selection criterion and claim to find six $z>$5 AGN in the CDF-S. Instead, Weigel et al. (2015) do not find any AGN at $z>$5 in the CDF-S, and Cappelluti et al. (2016 submitted) find four $z>$5 AGN in the CDF-S, only one of which was also in the Giallongo et al. (2015) sample.

\begin{table*}[h] 
\centering
\begin{tabular}{c|cccc|ccc|ccc|ccc}
\hline
\hline
Redshift & \multicolumn{4}{c}{Total}  & \multicolumn{3}{c}{z-spec} & \multicolumn{3}{c}{z-phot$_{peak}$} & \multicolumn{3}{c}{z-phot$_{w}$} \\ 
\hline 
    & \textbf{Total} & S  & H     & F       & S & H & F& S  & H & F  & S      & H    & F\\
$z$$>$3 & 190.2 & 145.1 & 13.2 & 31.9 & 77 & 4 & 6 & 58 & 8 & 21 & 68.1 & 9.2 & 25.9\\ 
$z$$>$4 & 30.6  &  21.2   & 2.0   & 7.4   & 10 & 0 & 1 & 10 & 2 & 4   & 11.2 & 2.0 & 6.4\\ 
$z$$>$5 & 8.6    &  5.9     & 0.4   & 2.3   & 2   & 0 & 0 & 4   & 1 & 2   & 3.9  &  0.4 & 2.3\\  
$z$$>$6 & 2.6    &  2.1    &  0.1   & 0.4   & 0   & 0 & 0 & 3   & 0 & 1   & 2.1  &  0.1  & 0.4\\ 
\hline
Redshift & \multicolumn{4}{c}{Total}  & \multicolumn{3}{c}{z-spec} & \multicolumn{3}{c}{z-phot$_{peak}$} & \multicolumn{3}{c}{z-phot$_{w}$} \\ 
\hline 
      & \textbf{Total} & S       & H      & F      & S  & H & F & S  & H & F  & S      & H    & F\\
$z$$>$3 & 179.4  & 143.4 & 12.1 & 23.9 & 77 & 3 & 5 & 55 & 6 & 18 & 64.4 & 9.1 & 18.9\\
$z$$>$4 & 28.2    & 20.4   & 2.0   & 5.8   & 10 & 0 & 1 & 9   & 1 & 3   & 10.4 & 2.0 & 4.8\\
$z$$>$5 & 7.8      & 5.4     & 0.4   & 2.0   & 2   & 0 & 0 & 3   & 1 &  1  & 3.4   & 0.4 & 2.0\\
$z$$>$6 & 2.1      & 1.6     & 0.1   & 0.4   & 0   & 0 & 0 & 2   & 0 & 1   & 1.6   & 0.1 & 0.4\\
\hline
\hline
\end{tabular}\caption{Top: number of sources in the high-redshift sample, divided by X-ray band adopted in the computation of the space density. First we use the soft-band (S; 0.5-2 keV) information: if DET$\_$ML$_S$$<$10.8, we use the hard-band (H; 2-10 keV) one. If a source has  DET$\_$ML$<$10.8 in both S and H, we use the information from the full band (F; 0.5-10 keV). Bottom: same as top, but taking into account only those sources actually used in the computation of the space density, after the application of a cut in the flux limit (i.e., 3.5$\times$10$^{-16}$ erg s$^{-1}$ cm$^{-2}$ in the soft band, 2.3$\times$10$^{-15}$ erg s$^{-1}$ cm$^{-2}$ in the hard band and 1.4$\times$10$^{-15}$ erg s$^{-1}$ cm$^{-2}$ in the full band). z-phot$_{peak}$ is the number of sources in a given bin assuming the PDF peak value, while z-phot$_{w}$ is the effective weighted contribution from all the PDF elements. z-phot$_{peak}$ and z-phot$_{w}$ numbers are given only for those sources with no $z_{spec}$ available. The total number of sources is computed adding $z_{spec}$ and the weighted contribution of $z_{phot}$.}\label{tab:sample}
\end{table*}

\subsection{Optically unidentified sources}\label{sec:no_optical}
80 sources in the whole \cha \leg survey have no optical $i$-band counterpart, no redshift, spectroscopic or photometric, available, and lie inside the optical/IR field of view. We further analyzed these sources, because some of them could be obscured and/or high-redshift AGN (Koekemoer et al. 2004). 

We visually inspected both the X-ray and the optical/IR images, centered at the X-ray position, and we found that about 50\% of the sources lack of optical counterpart because of either ($i$) low quality optical imaging, or ($ii$) the source is close to a very bright object (star or extended galaxy) and it is therefore undetected. 

A fraction of objects with no optical/IR counterpart are also candidate X-ray spurious detections. In the whole \cha \leg survey between 15 and 20 sources are expected to be spurious at DET\_ML=10.8 (Civano et al. 2016). Most will lie among those sources with DET\_ML$<$15, close to the survey limit, DET\_ML=10.8 

43 sources in the \cha \leg catalog are reliable X-ray sources (DET\_ML$>$15) without an optical $i$-band counterpart, but with a $K$-band (26 sources) or 3.6 $\mu$m IRAC counterpart (26 sources), or with no counterpart at all (10 sources).

\begin{figure}
\centering
\includegraphics[width=0.5\textwidth]{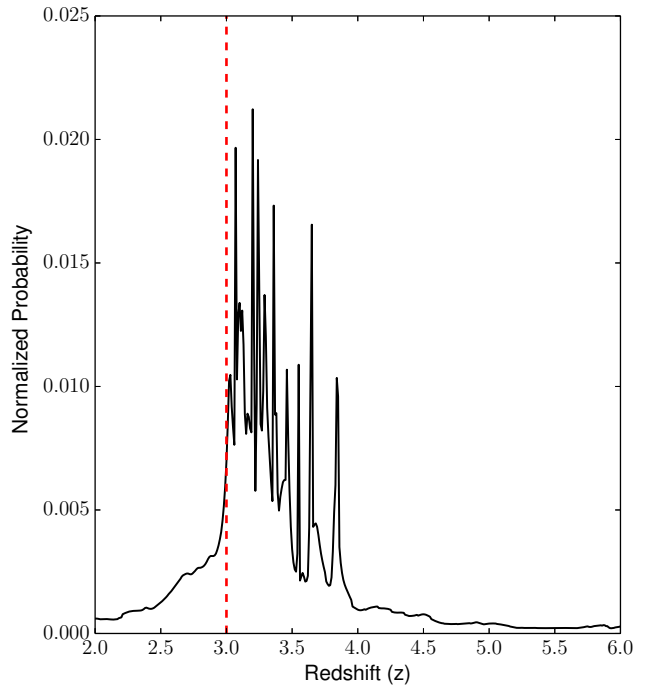}
\caption{Normalized probability distribution function of redshift for all sources with $z$ (either spectroscopic or photometric) $>$3: this distribution has also been used as redshift probability distribution function for the 43 sources in the sample without optical counterpart.}\label{fig:pdz_gt3}
\end{figure}

These 43 X-ray sources could be high-redshift candidates, or highly obscured sources, or both. We take all of them into account in the estimation of the upper boundary of our 2-10 keV space density (Section \ref{sec:space}), while we estimated the upper boundaries of the $z$$>$3 0.5-2 keV LogN-LogS using the 34 sources detected in the soft band. We assume that each of these sources has a PDF equal to the mean PDF of all the sources in L-COSMOS3 with z$\geq$3 (Figure \ref{fig:pdz_gt3}; the spikes in the distribution are associated to sources with $z_{spec}$, which usually have narrow PDFs). The contribution of the source in each bin of redshift has then been weighted by the value of the PDF at that redshift, as described above. 

\subsection{Optical/IR properties}\label{sec:opt_type}
\subsubsection{Photometry}
\begin{figure}
\centering
\includegraphics[width=0.5\textwidth]{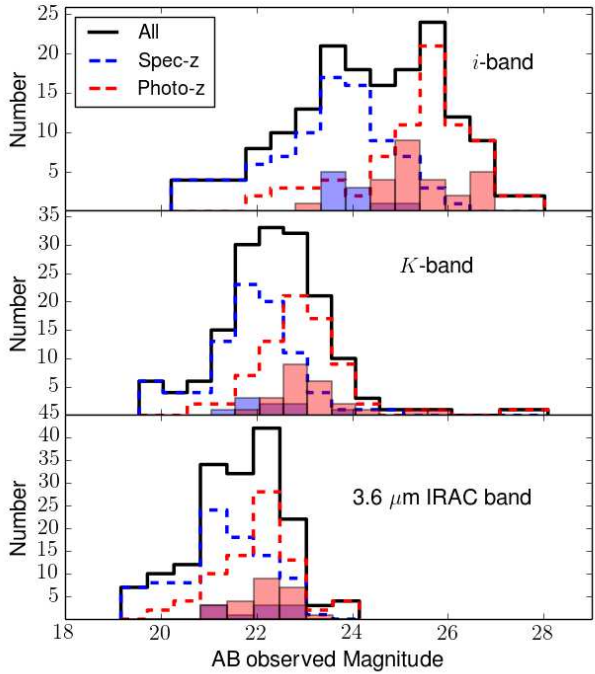}
\caption{Observed AB magnitude distribution in $i$- (top), $K$- and 3.6 $\mu$m IRAC-band (bottom) for the whole sample of sources with $z$$\geq$3 (black solid line), for the spectroscopic subsample (blue dashed line) and for the sources with only photometric redshift (red dashed line). Magnitude distribution for sources with no significant soft band emission is also shown in pale blue (spectroscopic subsample) and pale red (sources with only $z_{phot}$). Due to observational constraints, sources with spectroscopic redshift are also the optically and IR brightest ones.}\label{fig:histo_mag}
\end{figure}

In the L-COSMOS3 sample of 174 sources with $z\geq$3, 165 have $i$-band magnitude information (Capak et al. 2007, Ilbert et al. 2009, 2010; McCracken et al. 2010), 165 have $K$-band magnitude (Ilbert et al. 2013, Laigle et al. 2016) and 166 have 3.6 $\mu$m IRAC magnitude (Sanders et al. 2007, Laigle et al. 2016). The observed AB magnitude distributions in these three bands are shown in Figure \ref{fig:histo_mag}, dividing the sample in sources with spectroscopic redshift (blue dashed line) and with photometric redshift only (red dashed line) sources. Mean magnitudes in each band, for both sources with $z_{spec}$ and with $z_{phot}$ only, are shown in Table \ref{tab:photometry}.

Sources with spectroscopic redshift have average optical magnitude $\sim$2 dex brighter than sources with photometric redshift only. This is not a surprising result, since there is an inverse relation between the time required to obtain a reliable spectrum and the source brightness. Consequently, the $z_{phot}$ sub-sample covers AGN at $z>$3 with lower rest-frame near-ultraviolet (near-UV, 1000-3000 \AA) and optical (3000--6000 \AA) luminosities, which are observed in the $i$-band. 

The difference in magnitude is still significant, but smaller ($\sim$1 dex), in $K$-band, while  in the 3.6 $\mu$m IRAC band the difference is 0.7 dex. The $z_{spec}$ and $z_{phot}$ samples have similar magnitude distributions at longer wavelengths ($\simeq$6500--9000 \AA\ in the rest-frame, observed in the $K$-band). These objects could therefore be intrinsically fainter or more obscured than those for which we can provide a $z_{spec}$.

\begin{table}[H]
\centering
\begin{tabular}{ccccc}
\hline
\hline
Band & n$_{src}$ & mag$_{spec}$ & mag$_{phot}$ & mag$_{all}$\\
\hline
$i$ &165 & 23.4 & 25.3 & 24.3\\
$K$ & 165 & 21.9 & 23.0 & 22.4\\
3.6 $\mu$m IRAC & 166 & 21.3 & 22.0 & 21.7\\
\hline
\hline
\end{tabular}\caption{Number of sources and mean magnitude of sources with spectroscopic redshift, photometric redshift only and for the whole L-COSMOS3 sample, for each of the three optical/IR bands used in the \cha \leg counterpart detection procedure.}\label{tab:photometry}
\end{table}

\subsubsection{Spectroscopic and SED template types}
For most of the sources with an optical spectrum, we were able to determine the spectroscopic type of the AGN on the basis of the measured full width at half-maximum (FWHM) of the permitted emission lines. If one or more of these lines have FWHM$>$1000 km s$^{-1}$, we classify them as optically broad-line AGN (BLAGN; e.g., Vanden Berk et al. 2006; Stern \& Laor 2012), while sources with only narrow emission lines, or with only absorption lines, have been classified as non broad-line AGN (non-BLAGN).

Of the 87 sources with spectra, 54 are classified as BLAGN, while 28 are classified as non-BLAGN. For the remaining 5 sources, the spectral signal-to-noise ratio is not sufficiently high to draw safe conclusions on the presence or absence of broad lines. The mean $i$-band magnitude is 1.4 mag brighter for BLAGN ($\langle i_{AB}\rangle$=22.8) than for non-BLAGN ($\langle i_{AB}\rangle$=24.2).

For all the 174 L-COSMOS3 sources we also used an optical classification based on the best fit of the SED, computed with the publicly available code LePhare (Arnouts et al. 1999, Ilbert et al. 2006), based on $\chi^2$ template-fitting procedure, which is fully described in Salvato et al. 2011 (particularly, a summary of the template selection procedure is shown in their Figure 6). For 92 out of 174 sources, i.e., those with no spectral type, this is also the only type information available.

Based on the characteristic of the template best fitting the data, all the sources are divided in unobscured AGN, obscured AGN and galaxies. In the L-COSMOS3 sample with no spectral type information, 31 of the 92 sources are best fitted with an unobscured AGN template, 2 with an obscured AGN template and the remaining 59 with a galaxy template. It is worth noticing that all the L-COSMOS3 sources have L$_{\rm 2-10keV}>$10$^{43}$ erg $s^{-1}$ and are therefore AGN. The predominance of galaxy template best-fitted sources is mainly due to the procedure used to determine the templates used in the fit: all extended sources with flux in the 0.5-2 keV band $f_{0.5-2}$$<$8$\times$10$^{-15}$ erg s$^{-1}$ cm$^{-2}$ are fitted with a galaxy template, which best reproduces the SED of these usually optically faint galaxy-dominated sources (Salvato et al. 2011). 

Once again, the mean $i$-band magnitude is brighter for unobscured ($\langle i_{AB}\rangle$=24.3) than for obscured sources ($\langle i_{AB}\rangle$=25.7). 

For the 82 sources with spectral types the agreement between the spectral and the photometric classification is very good: 85\% of spectroscopic BLAGN are best--fitted with an unobscured AGN template, while 79\% of the spectroscopic non-BLAGN are best--fitted with an obscured AGN template or a galaxy template. 

The first difference can be explained with the fact that low-luminosity BLAGN SEDs can be contaminated by stellar light (Luo et al. 2010; Elvis et al. 2012; Hao et al. 2014; but see also Pons \& Watson 2014 on ``elusive-AGN'').  The latter discrepancy can instead be caused by the low quality of some of the spectra in L-COSMOS3, which can provide a reliable redshift but a less safe estimate of the presence of a broad line. It is also worth noticing that the spectroscopic classification is based on the presence of at least one broad line on a wavelength range of about 5000 \AA, while the SED classification is based on the minimum $\chi^2$ computed on a much larger bandwidth, from UV to NIR.

\subsubsection{Summary}
For the remaining part of our analysis, and especially in the analysis of the space density for obscured and unobscured sources (Section \ref{sec:space_type}), we divide our sample in Type 1, unobscured sources, and Type 2, obscured sources.
\begin{enumerate}
\item L-COSMOS3 contains 85 unobscured Type 1 AGN (49\% of the whole sample): 54 of these sources are spectroscopically classified BL-AGN, the remaining 31 are sources with no spectral type and fitted with an unobscured AGN SED template. 
\item L-COSMOS3 contains 89 obscured Type 2 AGN (51\% of the whole sample). We include in this sample the 28 spectroscopically classified non-BLAGN, the two sources best-fitted with an obscured AGN template, and the 59 best-fitted with a galaxy template.
\end{enumerate}

\begin{figure}
\centering
\includegraphics[width=0.5\textwidth]{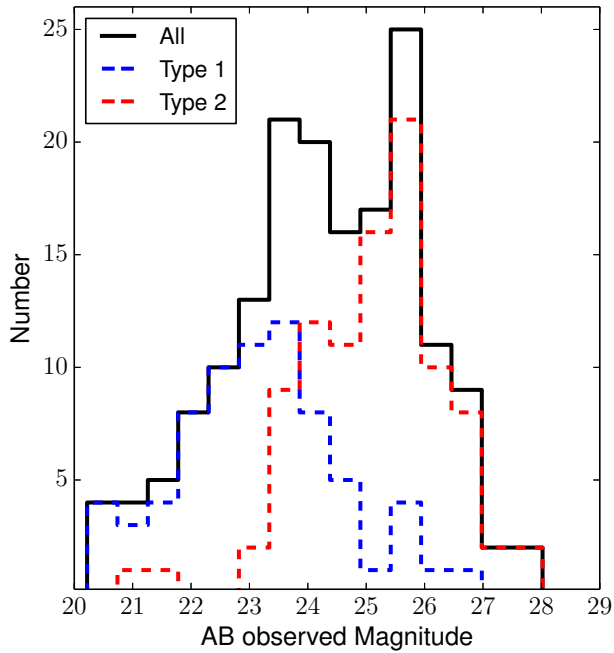}
\caption{Observed AB magnitude distribution in $i$-band for the whole sample of sources with $z$$\geq$3 (black solid line), for Type 1 AGN (blue dashed line) and for Type 2 AGN (red dashed line).}\label{fig:histo_mag_type}
\end{figure}

We show in Figure \ref{fig:histo_mag_type} the observed  $i$-band AB magnitude distribution for all the sources with $z$$\geq$3 (black solid line), for Type 1 AGN (blue dashed line) and for Type 2 AGN (red dashed line). The mean (median) $i$-band magnitude is 23.4 (23.4) for Type 1 AGN and 25.3 (25.4) for Type 2 AGN. The hypothesis that the two magnitude distributions are derived by the same parent distribution is rejected by a KS test, with a p-value$\simeq$1.7$\times$10$^{-14}$. 

The majority (66\%) of the sources with spectroscopic information are BLAGN, thus being brighter in $i$-band (see Figure \ref{fig:histo_mag}), which at the mean redshift of our distribution ($z\sim$3.5) samples the so called ``big blue bump'', emitting in the rest-frame UV (e.g., Shields 1978; Malkan \& Sargent 1982). The majority (66\%) of sources with only photometric information also have SED best-fitted with an obscured AGN or a galaxy template, which is consistent with the fact that these sources are intrinsically redder and thus fainter in the $i$-band (see also Brusa et al. 2010, Lanzuisi et al. 2013).

\section{0.5-2 keV AGN number counts}\label{sec:logn-logs}
We produced the high-z LogN-LogS relation, i.e., the number of sources $N$($> S$) per square degree at fluxes brighter than a given flux $S$ (erg s$^{-1}$ cm$^{-2}$). In our analysis, we treated our photometric redshifts as a sum of PDF contributions (see Section \ref{sec:photo-z}).
We derived the LogN-LogS in the observed soft band at $z$$>$3 and $z$$>$4. Recall that at these redshifts the 0.5-2 keV band roughly corresponds to the 2-10 keV rest-frame band. For the first time, we have a sample large enough to put constraints on the number counts also at $z$$>$5 (7.8 effective objects) and even $z$$>$6 (2.1 effective objects). 

The number counts were derived by folding our flux distribution through the sky coverage (i.e. the area of the survey covered at a given flux) of the \cha \leg survey (Civano et al. 2016).

The sensitivity curve which describes the sky coverage is very steep in the flux regime close to the flux limit of the survey, leading to uncertainties on the area that are larger than at bright fluxes. To avoid these uncertainties, and to reduce the Eddington bias, we applied a cut in flux corresponding to 10\% of the total area of the survey. Hence, we took into account only sources with a 0.5-2 keV flux above  3.5$\times$ 10$^{-16}$ erg s$^{-1}$ cm$^{-2}$. The sample used for the number counts therefore includes 143.4 effective sources at $z$$>$3, 20.4 at $z$$>$4, 5.4 at $z$$>$5 and 1.6 at $z$$>$6.

We computed the cumulative source distribution with the equation:

\begin{equation}
N(>S)=\sum_{i=1}^{N_S}\frac{w_i}{\Omega_i} [deg^{-2}],
\end{equation}

where $N(>S)$ is the number of surces with flux greater than a given flux $S$, $\Omega_i$ is the sky coverage associated to the flux of the $i$th source, $N_S$ is the number of sources above flux $S$ and $w_{i}$ is the weight linked to the PDF contribution, $w_{i}$=$\frac{PDF(z)}{\sum_0^7{PDF(z)}}$ ($w_{i}$=1 for sources with a spectroscopic redshift). 
We computed the 90\% uncertainties on the number counts using the Bootstrap technique. We first randomly resampled 10,000 times the original input source list, obtaining 10,000 new lists of sources having the same size of the original one; we then computed the LogN-LogS for each of these resamples and the 5th and 95th percentiles of the 10,000 number counts, in each bin of flux.

\begin{figure*}
  \begin{minipage}[b]{.5\linewidth}
    \centering
  \includegraphics[width=1.02\linewidth]{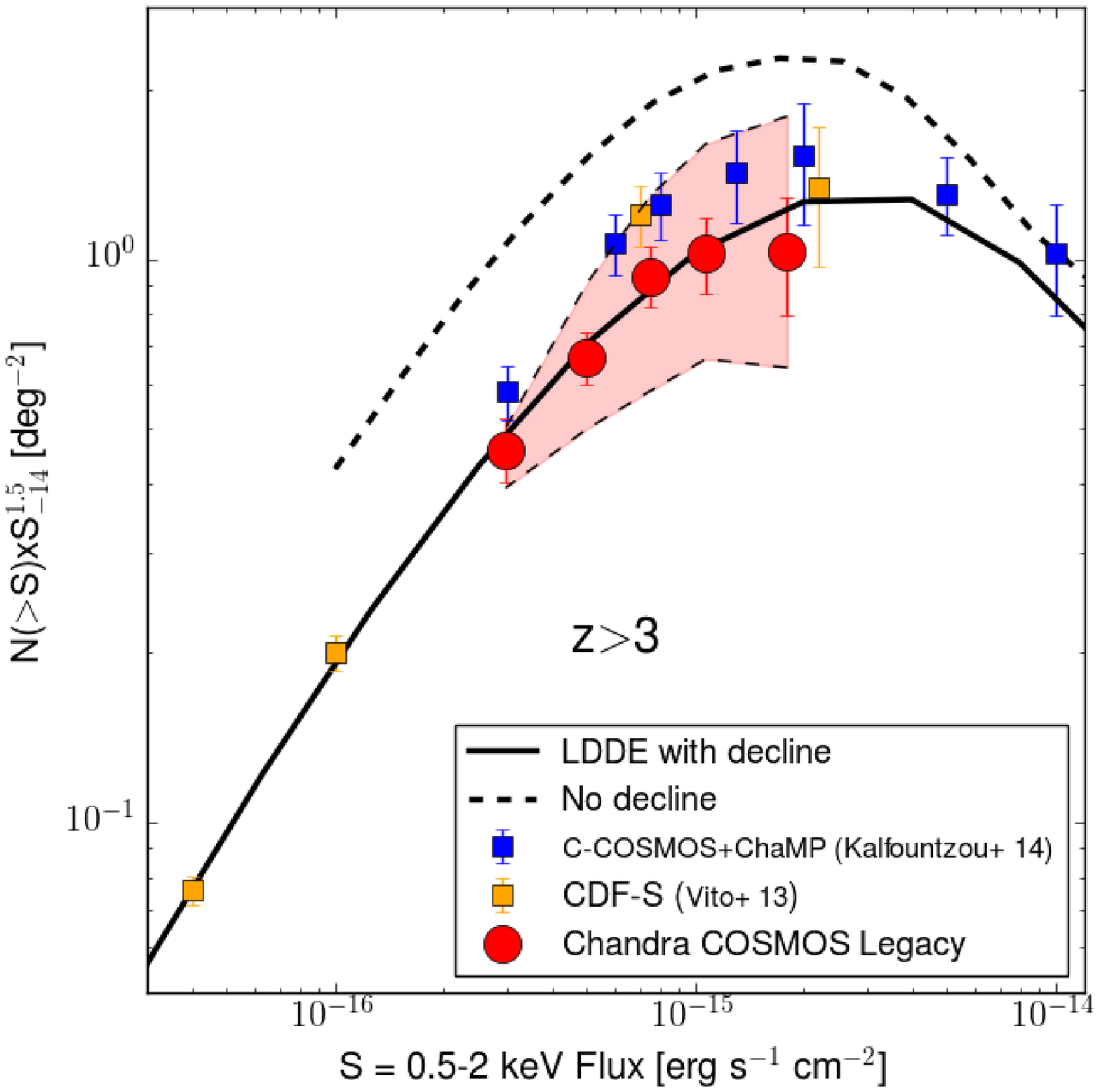}
  \end{minipage}
  \hfill
  \begin{minipage}[b]{.5\linewidth}
    \centering
  \includegraphics[width=.98\linewidth]{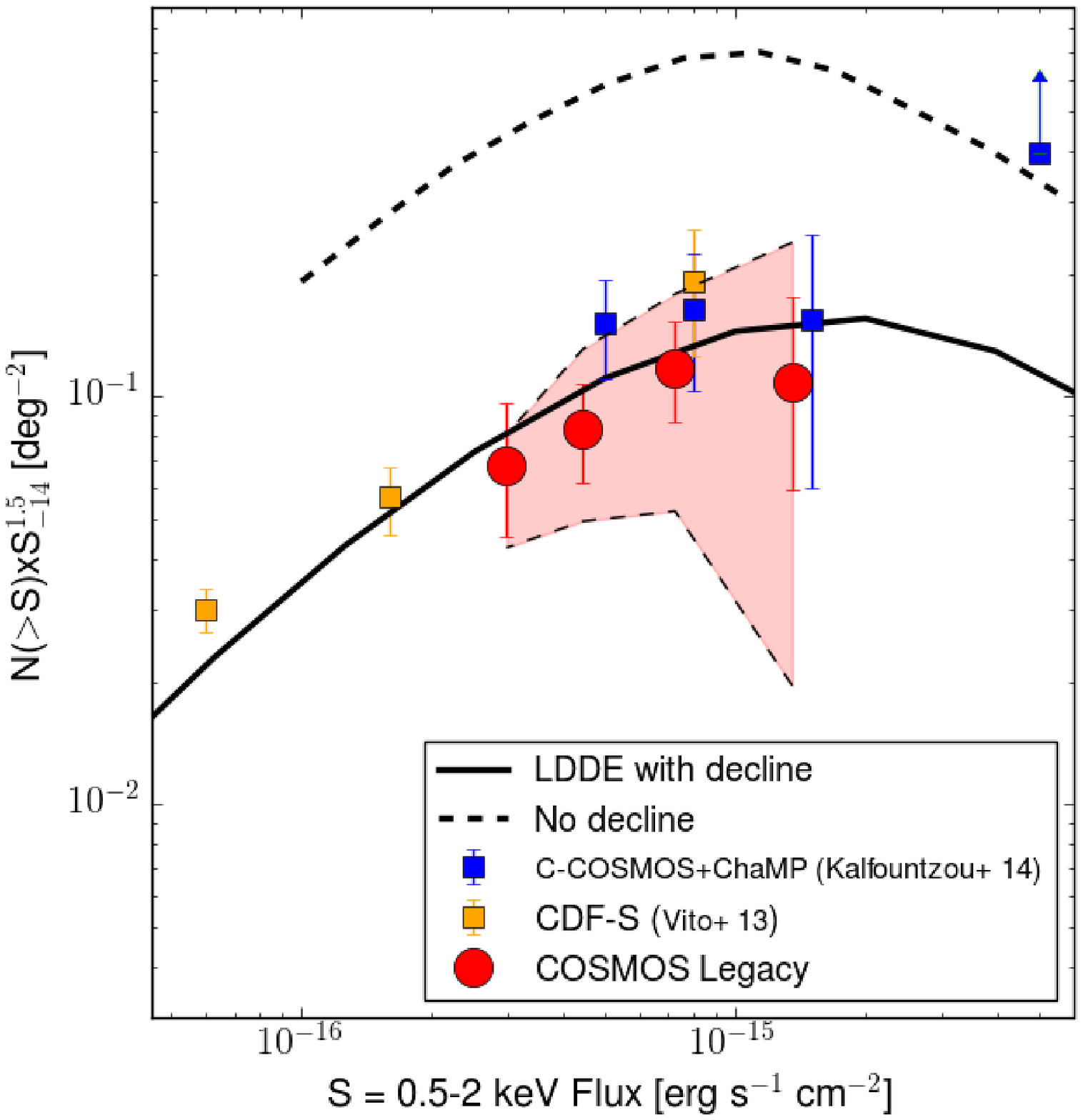}
\end{minipage}
\caption{Euclidean normalized LogN-LogS relation in the 0.5-2 keV band for \cha \leg (red circles), for $z$$>$3 (left) and $z$$>$4 (right). Results from Vito et al. (2013, 4 Ms CDF-S, orange squares), and Kalfountzou et al. (2014, C-COSMOS and ChaMP data, blue squares)  are also shown for comparison, together with examples of models with (black solid line, from Gilli et al. 2007) and without exponential decline (dashed line, from Treister et al. 2009). The pale red area is obtained computing the number counts adding and subtracting to the flux value its 1$\sigma$ error. In the computation of the upper boundary the weighted contribution of sources with no optical counterpart is also taken into account. All number counts are multiplied by $(S /10^{14})^{1.5}$ to highlight the deviations from the Euclidean behavior.}\label{fig:logn-logs_1}
\end{figure*}

\begin{figure}
\centering
\includegraphics[width=0.5\textwidth]{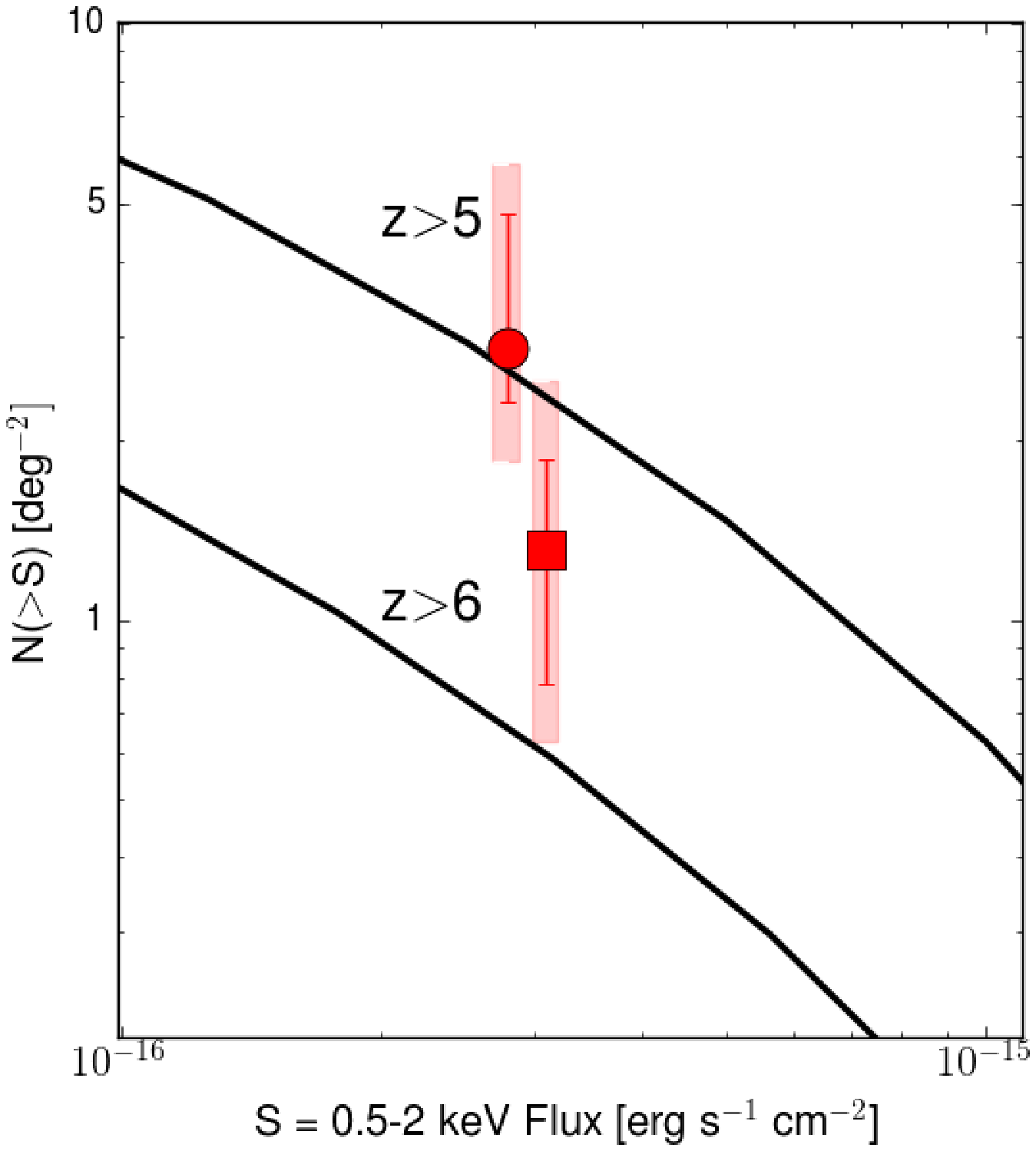}
\caption{LogN-LogS relation in the 0.5-2 keV band for \cha \leg (red circles), for $z$$>$5 (upper part) and $z$$>$6 (lower part). Models from Gilli et al. (2007, black solid line) are also shown for comparison. The pale red area is obtained computing the number counts adding and subtracting to the flux value its 1$\sigma$ error. In the computation of the upper boundary the weighted contribution of sources with no optical counterpart is also taken into account.}\label{fig:logn-logs_2}
\end{figure}

We show our euclidean normalized LogN-LogS relations (i.e., with $N(>S)$ multiplied by $S^{1.5}$) in Figures \ref{fig:logn-logs_1} ($z$$>$3, left, and $z$$>$4, right, red circles) and \ref{fig:logn-logs_2} ($z$$>$5, left, and $z$$>$6, right). We also estimated upper and lower boundaries of the logN-logS (plotted as the black dashed lines limiting the pale red area), as follows:
\begin{enumerate}
\item for the upper boundary we computed $\Omega_i$ for each source adding to the observed flux the 1$\sigma$ uncertainty on the flux, and we added to the sample also the 34 soft emitting sources with no optical counterpart, assuming for each of them a PDF equal to the average PDF of sources with $z$$>$3 (see Section \ref{sec:no_optical}). With this second addition, we are under the strong assumption that all the non-detections in the optical bands are actually high-redshift X-ray selected sources;
\item for the lower boundary we computed $\Omega_i$ for each source after subtracting the 1$\sigma$ uncertainty on the flux to the observed flux.
\end{enumerate} 

In Figure \ref{fig:logn-logs_1} we also plot the Euclidean normalized number counts from two other studies: the Vito et al. (2013), using 4-Ms \cha Deep Field-South data (yellow squares), and the Kalfountzou et al. (2014), using C-COSMOS and ChaMP data (orange squares). These studies used datasets that cover the range from deep, pencil-beam area (CDF-S, 0.13 deg$^2$, flux limit in the 0.5-2 keV band $f_X$$\simeq$ 9.1$\times$ 10$^{-18}$ erg s$^{-1}$ cm$^{-2}$, Xue et al. 2011), 
to large areas and intermediate depth, combining C-COSMOS (0.9 deg$^2$, flux limit in the 0.5-2 keV band $f_X$$\simeq$ 1.9$\times$ 10$^{-16}$ erg s$^{-1}$ cm$^{-2}$, Elvis et al. 2009) and the non-contiguous field ChaMP ($\simeq$30 deg$^2$, flux limit in the 0.5-2 keV band 3.7$\times$ 10$^{-16}$ erg s$^{-1}$ cm$^{-2}$, Kim et al. 2007; Green et al. 2009). 

The L-COSMOS3 results are in general agreement with these two other studies, both at bright and faint fluxes, but with a significant improvement in the uncertainties. At $z>$3, the 90\% confidence error-bars for L-COSMOS3 are 20--40\% smaller than the Poissonian uncertainties measured by the other studies. At $z$$>$4, 
the L-COSMOS3 number counts normalization is slightly lower, but consistent within the uncertainties, with those in Kalfountzou et al. (2014) at $f_X$$<$5$\times$ 10$^{-16}$ erg s$^{-1}$ cm$^{-2}$. The L-COSMOS3 data also show a declining trend consistent with the results from Vito et al. (2013) at $f_X$$<$3 $\times$ 10$^{-16}$ erg s$^{-1}$ cm$^{-2}$, that was not present in Kalfountzou et al. (2014). 

Due to our good statistics, we are able to improve the constraints on predictions of different phenomenological models. We show the different model predictions in Figures \ref{fig:logn-logs_1} and \ref{fig:logn-logs_2}, as black curves. We do not show the predictions of the FDPL model because Aird et al. (2015) computed only 2-10 luminosity functions and space density: we will discuss their predictions in Section \ref{sec:space}. We also do not compare our number counts with physical models, but we will discuss the space density predicted by one of this models (Shen 2009) in Section \ref{sec:merger}, in comparison with the L-COSMOS3 space density.
\begin{enumerate}
\item The thick solid lines show the predictions of an X-ray background (XRB) synthesis model with high-redshift exponential decline. The model we show is the Gilli et al. (2007) one, based on the extrapolation of the X-ray luminosity function observed in a low-redshift regime (Hasinger et al. 2005) and parametrized with an LDDE model and with a high-redshift exponential decline, as in Schmidt, Schneider \& Gunn (1995): $\Phi$(z)=$\Phi$(z$_0$)$\times$10$^{-0.43(z-z_0)}$ (with z$_0$=2.7). This model was developed in order to fit the optical luminosity function in the redshift range z=[2.5-6] (Fan et al. 2001). 
\item An example of model without exponential decline is shown as a dashed line. The model we show is the X-ray background population synthesis model by Treister, Urry \& Virani (2009), and is based on the luminosity function estimated by Ueda et al. (2003). 
\end{enumerate}

At $z$$>$3 (Figure \ref{fig:logn-logs_1}, left) our results indicate that a decline in the number of counts is needed in the X-ray as well as in in the optical band. The predictions of the model with no exponential decline are too high by a factor $\simeq$2 in comparison to our data, at any flux. This result is not fully unexpected, and it has been already observed in previous works (see, e.g., Civano et al. 2011; Vito et al. 2013; Kalfountzou et al. 2014). 

The LDDEexp model predictions fit the L-COSMOS3 results, within the 1 $\sigma$ uncertainties, also at $z$$>$4 (Figure \ref{fig:logn-logs_1}, right). This improve the results reported in Kalfountzou et al. (2014), which also showed a good agreement between the data and the LDDEexp, although with larger uncertainties.

In Figure \ref{fig:logn-logs_2} we show the first analysis ever of X-ray selected AGN number counts at $z$$>$5 (upper part) and $z$$>$6 (lower part). At $z$$>$5, our data (red circle) are in agreement with the LDDEexp model (solid line). 
At $z$$>$6 our data (red square) are slightly above the predictions of the LDDEexp model (solid line).
In both panels, we do not show the predictions of the model without exponential decline because we already ruled them out in the $z>$3 and $z>$4 analysis.%

\section{2-10 keV comoving space density}\label{sec:space}
For the computation of the space density in the 2-10 keV band, we applied the flux cuts described in Section \ref{sec:sample_summary} to avoid the Eddington bias at faint fluxes. the fluxes at which these cuts are applied are 3.5$\times$10$^{-16}$ erg s$^{-1}$ cm$^{-2}$ in the soft band, 2.3$\times$10$^{-15}$ erg s$^{-1}$ cm$^{-2}$ in the hard band and 1.4$\times$10$^{-15}$ erg s$^{-1}$ cm$^{-2}$ in the full band. 
We report a summary of the final number of sources included in the space density sample in Table \ref{tab:sample} (bottom). As can be seen, more than 80\% of the sources in the sample are detected in the 0.5-2 keV observed band (i.e., the band that at $z>$3 roughly corresponds to the 2-10 keV rest-frame band). However, to complete our analysis we computed the extrapolated 2-10 keV rest-frame luminosity also for those sources with no significant 0.5-2 keV detection, using first the 2-10 keV observed flux and, for those sources with no significant 2-10 keV detection, the 0.5-10 keV observed flux. The fluxes and luminosities are estimated assuming $\Gamma$=1.4, the X-ray background slope and therefore a good average slope for a population of both obscured and unobscured AGN (e.g., Markevitch et al. 2003). 

We computed the comoving space density using the 1/$V_{Max}$ method (Schmidt 1968), corrected to take into account the fact that in our survey the area is flux dependent. We also worked with the assumptions described in Avni \& Bahcall (1980), which take into account the fact that each source could in principle have been found at any X-ray depth within the survey limits.

For every redshift associated to a source in L-COSMOS3, spectroscopic or photometric with an associated PDF($z_{bin}$) $>$0 in at least one bin of redshift $z_{bin}$$\geq$3, we computed the maximum available volume over which the source can be detected, using the equation

\begin{equation}
V_{max}=\int_{z_{min}}^{z_{max}} \! w(z) \Omega_{band}(f(L_{\rm X}, z)) \frac{\mathrm{d}V}{\mathrm{d}z}\, \mathrm{d}z,
\end{equation}

where $w$ is the weight linked to the PDF contribution, $w$=$\frac{PDF(z)}{\sum_0^7{PDF(z)}}$ ($w$=1 for sources with a spectroscopic redshift), $\Omega_{band}$(f($L_{\rm X}$, $z$)) is the sky coverage at the flux $f$($L_{\rm X}$, $z$) observed from a source with redshift $z$ and intrinsic luminosity $L_{\rm X}$, in the band where the flux was estimated; $z_{min}$ is the lower value of the redshift bin and $z_{max}$ is the minimum value between the maximum observable redshift of the source at the flux limit of the survey and $z_{up,bin}$, the upper value of the redshift bin. No absorption correction is applied to the fluxes: however, while estimating the obscuration correction from the X-ray hardness ratio for the whole \cha \leg sample (Marchesi et al. 2016) we found that in the 2-10 keV band the correction is larger than 20\% for less than 10\% of the sources, and is always smaller than 50\%. We used the flux $f$ from the first available band where DET\_ML$>$10.8, starting from 0.5-2 keV, then 2-10 keV and finally 0.5-10 keV. The flux was then converted to the 2-10 keV luminosity, using the equation

\begingroup\makeatletter\def\f@size{9}\check@mathfonts
\begin{equation}
L_{2-10 keV,rest}= \frac{4\pi d_l(z)^2 \times f \times (10^{2-\Gamma}-2^{2-\Gamma} )}{(E_{max}(1+z))^{2-\Gamma}-(E_{min}(1+z))^{2-\Gamma}},
\end{equation}\endgroup

where E$_{min}$ and E$_{max}$ are the minimum and maximum energies in the range where the flux is measured, and $d_l(z)$ is the luminosity distance at the given redshift.

Finally, we summed the reciprocal of all V$_{max}$ values in each redshift bin [$z_{min}$-$z_{max}$] in order to compute the comoving space density value $\Phi$:

\begin{equation}
\Phi= \sum_{i=1}^{z_{min}<z<z_{max}}\left(\frac{1}{V_{max,i}}\right).
\end{equation}

The 90\% uncertainties on the space density values have been computed through the bootstrap technique, randomly resampling 10,000 times our list of sources, in the same way described in Section \ref{sec:logn-logs} for the number counts.

\begin{figure}
\centering
\includegraphics[width=0.5\textwidth]{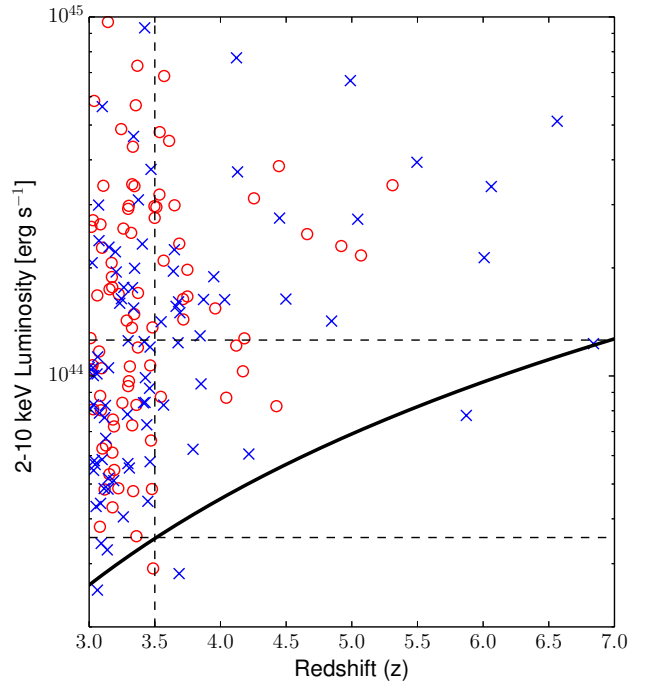}
\caption{Evolution with redshift of the 2-10 keV K-corrected luminosity for all sources in L-COSMOS3 with a spectroscopic (red circles) or a photometric (blue crosses) redshift. The black solid line shows the 10\%-area limit of the survey, computed from the 0.5--2 keV flux limit, while the black dashed lines highlight the two subsamples used in the computation of the space density (Log($L_{\rm X}$)=[43.55-44.1] over the redshift range $z$=[3-3.5] and Log($L_{\rm X}$)$>$44.1 over the redshift range $z$=[3-6.6]).}\label{fig:z_vs_lx}
\end{figure}

We divided L-COSMOS3 in two different luminosity ranges for completeness reasons (see Figure \ref{fig:z_vs_lx}). The high-luminosity space density has therefore been computed in six redshift bins in the range $z$=[3-6.6], with Log($L_{\rm X}$)$>$44.1. The low-luminosity space density, instead, has been computed in three redshift bins in the range $z$=[3-3.5], with 43.55$\leq$Log($L_{\rm X}$)$<$44.1\footnote{These luminosity ranges are slightly different from those adopted for C-COSMOS (Civano et al. 2011), where the low-luminosity range was Log($L_{\rm X}$)=[43.55-44.15] and the high-luminosity one was Log($L_{\rm X}$)$>$44.15. This difference is due to the fact that in Civano et al. (2011) a power-law with $\Gamma$=2 was adopted in the flux computation, while we use $\Gamma$=1.4.}. It is worth noticing that the last redshift bin of our high-luminosity space density ($z$=[5.5--6.6]) contains only photometric redshifts. As we explained in Section \ref{sec:sample_summary}, the photo-$z$ estimation at these extreme redshifts is based on significantly less photometric points ($<$10) than at lower redshifts. Therefore, while taking into account the PDF contribution at $z>$5.5 for the completeness of our analysis, we also claim that at these redshifts our results are subjected to significant uncertainties and need to be confirmed (e.g., with spectroscopic follow-ups of candidate $z>$6 sources).

To improve our analysis, we estimated upper and lower boundaries of the space densities, using as input parameters in the X-ray fluxes plus or minus their 1$\sigma$ uncertainties. 

In the computation of the upper boundary we also take into account the 43 sources with no optical counterpart. As explained in Section \ref{sec:no_optical}, these sources are candidate high-redshift AGN. For each source, we assumed as PDF the mean PDF for all the sources in L-COSMOS3 with $z\geq$3 (Figure \ref{fig:pdz_gt3}). We then computed the space density for this subsample with the same technique described above. The values of $\Phi$ that we obtained have then been summed to the upper boundary obtained using $f_X$+$\sigma$($f_X$) as input parameter.

\subsection{Log($L_{\rm X}$)$>$44.1 space density}
The L-COSMOS3 space density at Log($L_{\rm X}$)$>$44.1 is shown in Figure \ref{fig:space1}, left panel (red dots). The best linear fit to our data ($\Phi$=$a$+$bz$, red solid line), has a slope $b$=--0.46$\pm$0.04. We observe a decline of a factor $\sim$20 in the space density from $z$=3 to $z$=6.2. It is interesting to note that Trakhtenbrot et al. (2015b, submitted) measured the black hole masses and accretion rates of a sample of 10 L-COSMOS3 sources at $z\sim$3.3, and, on the basis of their results, estimated that a large population of $z>$5 AGN with $M_{\rm BH}\sim$ 10$^{6-7}$ M$_\odot$ and $L_{\rm 2-10keV}\geq$10$^{43}$ erg s$^{-1}$ should exist and be observable. The lack of this population in our dataset can be due to increased obscuration at $z>$5 with respect to $z\sim$3, or to a lower radiative efficiency in the early phase of black hole growth.

We compare our space density with the one of Vito et al. (2014, orange squares). We remind that the results of this work are best-fitted by a PDE model. There is a generally good agreement between their results and ours, at all redshifts. It is also worth noticing that the work of Vito et al. (2014) is based on several assumptions that differ from those used in this work, e.g., they assume a photon index $\Gamma$=1.8--1.9\footnote{Vito et al. (2014) space density is obtained combining sources from different surveys, having different assumptions on the rate-to-flux conversion procedure.}; moreover, they use photometric redshifts without weighting the PDF contribution.

We also compared our results with the predictions from the LDDEexp models from Gilli et al. (2007, black solid line), Ueda et al. (2014, cyan dashed line), Miyaji et al. (2015, green solid line), and with those from the FDPL model of Aird et al. (2015, black dashed line). We described the Gilli et al. (2007) in Section \ref{sec:logn-logs}. The Ueda et al. (2014) and the Miyaji et al. (2015) models are both derivations of the LDDE model, while the FDPL model has been derived independently.

The FDPL model is higher than our data by a factor 2 at 3$<$$z$$<$5, even if the upper boundaries are considered, at high luminosities (Figure \ref{fig:space1}, left). Our data are in better agreement with the predictions of the various LDDEexp models, with a discrepancy by a factor smaller than 2 in the redshift range $z$=[3-4], while at higher redshift the predictions of the model are in agreement with our data. There is a good agreement between the slope of our space density ($b$=--0.45$\pm$0.02), and the one of the different LDDE models (e.g., the Gilli et al. 2007 model slope is $b$=-0.53). However, we point out that the models are based on several different assumptions, and some of them differ by the one we use in this work. For example, we assume a fixed photon index $\Gamma$=1.4 to compute the rate-to-flux (and therefore luminosity) conversion factors, while the FDPL space density is computed assuming a distribution of different photon indexes. Moreover, the models X-ray luminosities are absorption-corrected, while those in our work are ``observed'' luminosities, since the majority of the L-COSMOS3 sources does not have a photon statistics good enough to properly compute the absorption contribution.

We also show results from optical surveys such as Masters et al.  (2012, black diamonds, left; sample of Type 1 objects only), McGreer et al. (2013, blue diamonds, left; sample of Type 1 objects only), Ikeda et al. (2011, black diamonds, right) and Glikman et al. (2011, purple diamonds, right). It is worth noticing that the Masters et al. (2012) sample was obtained in the COSMOS field and is overlapping with L-COSMOS3. To compare the optical results to ours in  \cha \leg, we assumed the relation between the X-ray luminosity at 2 keV, L$_{2keV}$, and the luminosity at 1500 \AA, $L_{1500}$, from Young et al. (2010),

\begin{equation}
\alpha_{OX}=1.929-0.119\ log L_{1500},
\end{equation}

with 

\begin{equation}
\alpha_{\rm OX} = \frac{\log\left(L_{\rm 2 keV}/L_{1500}\right)}{\log\left(\nu_{\rm 2 keV}/\nu_{1500}\right)}
\end{equation}

We then integrated the luminosity functions of Masters et al. (2012) and McGreer et al. (2013) down to $M_{1450}$=--24.5, corresponding approximately to Log($L_{\rm X}$)$\sim$44.1, and we compared them with our high-luminosity space density.  The slope derived from the optical surveys ($b$=-0.68$\pm$0.02) is in good agreement with our data and with the different LDDE models; the normalization is instead $\sim$4--5 times lower in the optical space density than in the X-ray one, due to the large fraction of obscured sources that are not detected in the optical band.

\subsection{43.55$<$Log($L_{\rm X}$)$<$44.1 space density}
As in the high luminosity regime, also in the low luminosity one (Log$L_{\rm X}$=[43.55-44.1]) we observe a decline in the space density values moving toward higher redshifts. The best linear fit to our data ($\Phi$=$a$+$bz$, red solid line), has a slope $b$=--0.82$\pm$0.18. 

This result seems in slightly better agreement with the LDDEexp models than with the FDPL model, in the redshift range, $z$=[3--3.4]: in this redshift range, the FDPL model underpredicts with respect to our data by 60--80\%. We also find that our results are a factor $\sim$2--3 higher than those of Vito et al. (2014), although their data are affected by larger uncertainties than ours, due to the smaller size of their sample. 

\begin{figure*}
  \begin{minipage}[b]{.5\linewidth}
    \centering
  \includegraphics[width=.97\linewidth]{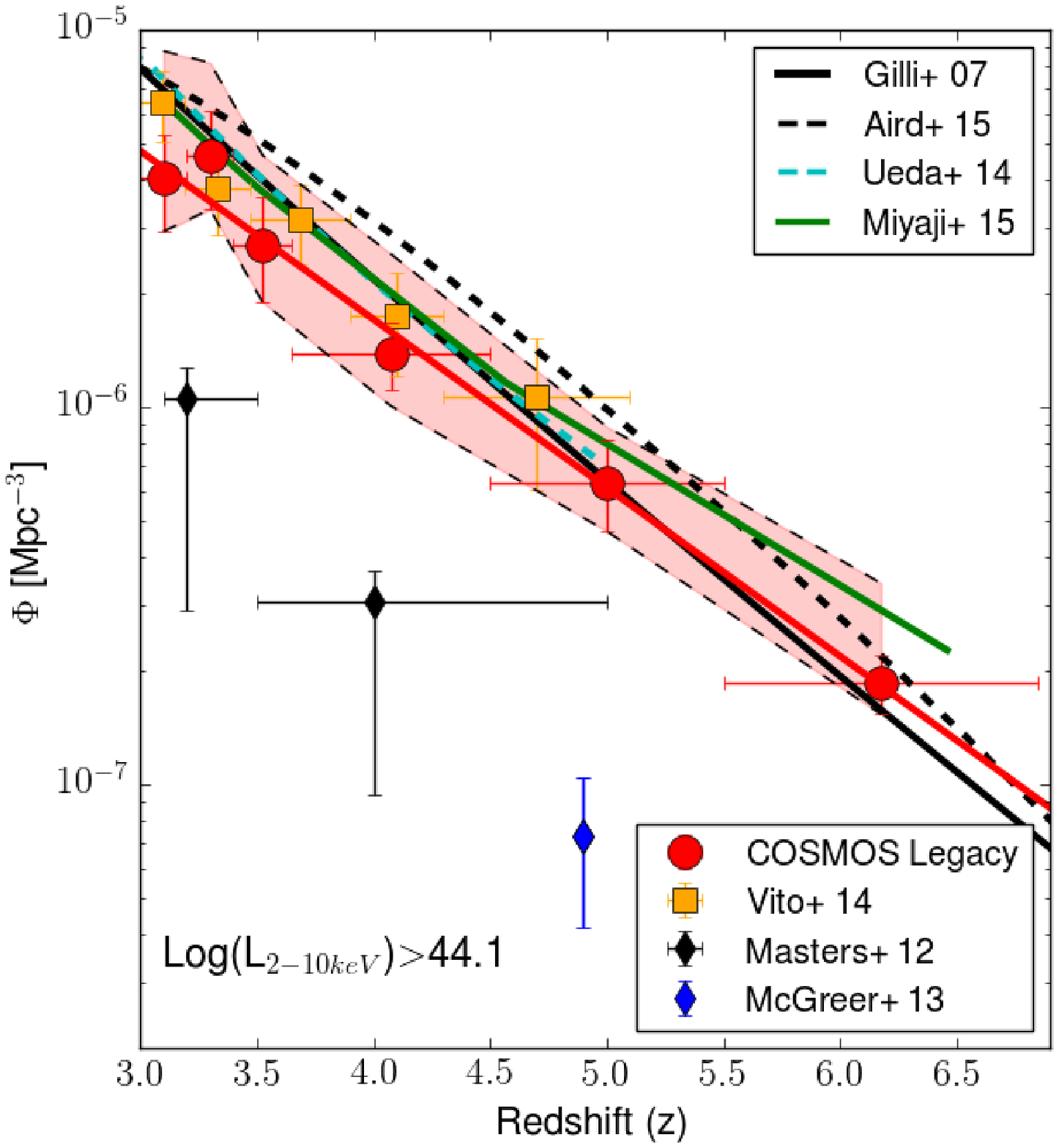}
\end{minipage}%
  \begin{minipage}[b]{.5\linewidth}
    \centering
  \includegraphics[width=.99\linewidth]{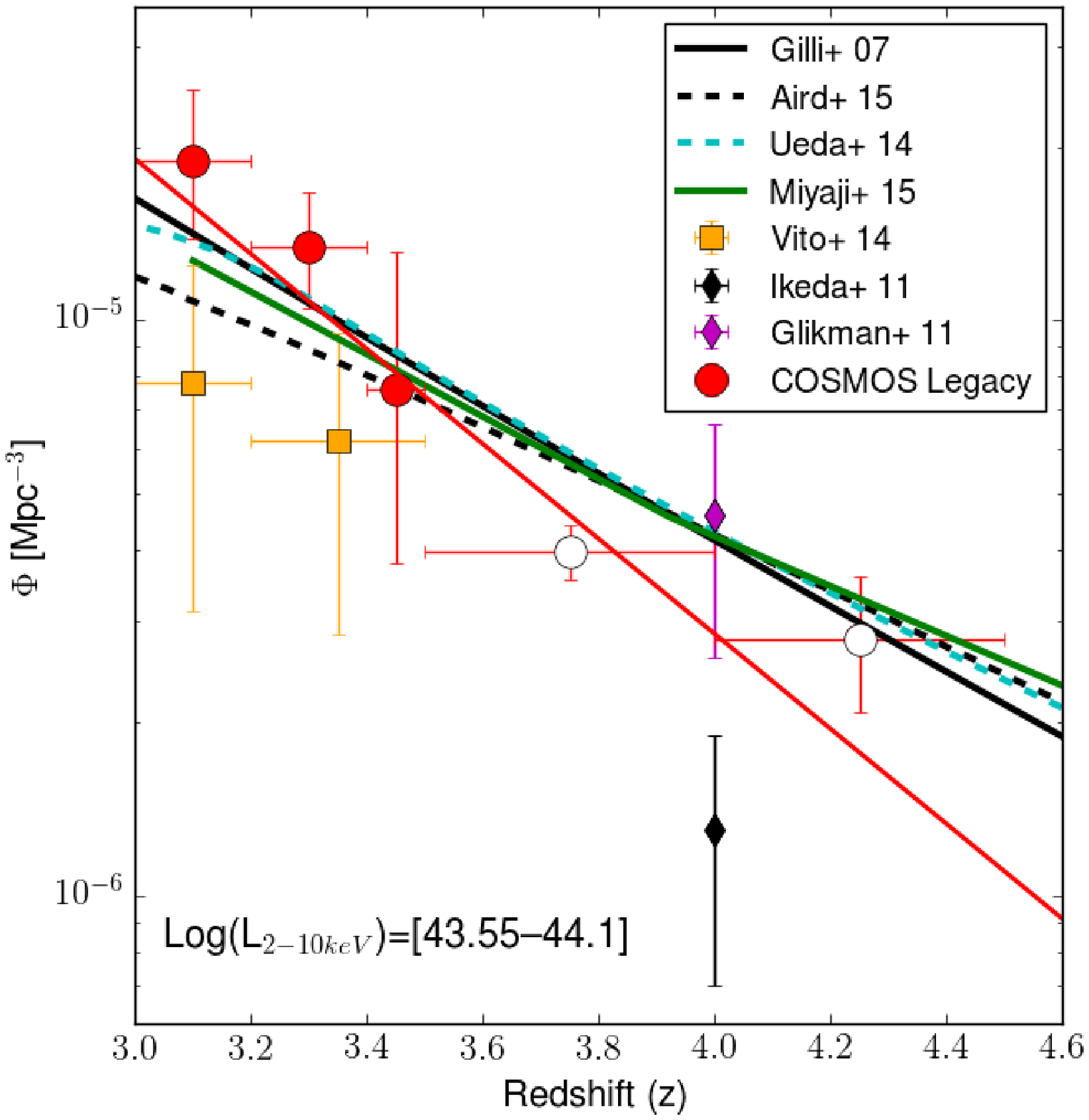}
\end{minipage}
\caption{Space density for sources with $L_{\rm X}$$>$10$^{44.1}$ (left) and 10$^{43.55}$$\leq$$L_{\rm X}$$<$10$^{44.1}$ (right), in the 2-10 keV band. The \cha \leg survey results are plotted with red dots, while results from Vito et al. (2014, orange squares) are also shown for comparison, together with optical space density from Masters et al. (2012, black diamonds) and McGreer et al. (2013, blue diamonds). Four different models of X-ray population synthesis are also shown, those of Gilli et al. (2007, black solid line), Aird et al. (2015, black dashed line), Ueda et al. (2014, cyan dashed line) and Miyaji et al. (2015, green solid line). The red solid line is the best fit to the \cha \leg data, assuming an equation $Log(\Phi)=a + b \times z$. The pale red area in the left figure is obtained computing the space density adding and subtracting to the flux value its 1$\sigma$ error. In the computation of the upper limit the weighted contribution of sources with no optical counterpart is also taken into account. The open markers in the right panel are the results obtained in those redshift bins where our survey is not complete, and therefore should be treated as lower limits.} 
\label{fig:space1}
\end{figure*}

In Figure \ref{fig:space1}, right panel, we also show the optical luminosity functions of Ikeda et al. (2011, black diamonds) and Glikman et al. (2011, purple diamonds): we integrated their luminosity functions in the absolute magnitude range $M_{1450}$=[-23.5;-21.8]. To compare these data at $z$=4 with our results, we computed the \cha \leg space density in two bins of redshift at $z$=[3.5--4.5], where L-COSMOS3 is not complete (see Figure \ref{fig:z_vs_lx}); therefore these two data-points should be treated as lower limits. We found that our data are in good agreement with the result obtained by Glikman et al. (2011), while the estimations by Ikeda et al. (2011) lie below our estimations by a factor of $\sim$2--3. It is however worth noticing that both these optical surveys are sampling unobscured Type 1 AGN, while in L-COSMOS3 is taken into account also a significant fraction of obscured objects. We will discuss the agreement between the optical surveys and our Type 1 AGN population space density in the next section.

\subsection{Obscured versus Unobscured AGN}\label{sec:space_type}
The high-redshift decline of space densities has been measured in both optical and X-ray selected AGN samples. Therefore, given that X-ray selected samples suffer considerably less obscuration bias compared to those optical selected, a similar trend should imply that the fraction of obscured AGN does not change significantly above z=3. In fact, previous works showed an increase in the fraction of obscured objects in the redshift range $z$=[1-2] (e.g., Ballantyne et al. 2006; Iwasawa et al. 2012), followed by a decline of this fraction at higher redshifts (Hasinger 2008; see also Gilli 2010 for a general review and an analysis of possible selection biases).  We test this result with L-COSMOS3, which we divide in two subsamples on the basis either of the spectroscopic classification (where available, i.e. for 82 sources) or the best fitting SED template (seeSection \ref{sec:opt_type} for further details). Summarizing, 85 sources with nominal redshift value $z$$\geq$3 are classified as Type 1, or unobscured, while the remaining 89 are classified as Type 2, or obscured. For the analysis of the space density, however, we also take into account (as for the general case) the weighted contribution of those sources with photometric redshift $z$$<$3 and PDF$>$0 in at least one bin of redshift with z$\geq$3. 

We point out that the optically-based classification of the source obscuration adopted here is less reliable than one based on a proper estimation of the intrinsic absorption ($N_{\rm H}$) based on the X-ray spectral fitting. For example, the template SED-fitting procedure can occasionally introduce biases, and a fraction of sources best-fitted by a galaxy SED template can be objects where the galaxy optical/IR contribution is dominant but no intrinsic absorption is present.
However, the X-ray spectral fitting requires at least few tens of counts in the 0.5--7 keV band (Lanzuisi et al. 2013), and only 20 out of 174 sources in L-COSMOS3 have such a number of counts. Moreover, at $z\geq$3 even the $N_{\rm H}$ estimate based on the source hardness ratio (HR=$\frac{H-S}{H+S}$, where H are the source net counts in the 2--7 keV band and S are the source net counts in the 0.5--2 keV band), which provides an estimate of the source intrinsic absorption at lower redshifts (see, e.g., Marchesi et al. 2016), is not reliable due to the higher degeneracy in the HR-$z$ space of objects with significantly different $N_{\rm H}$.
However, there are at least two evidences that suggest at least a fair agreement between the X-ray and the optical obscuration classification in L-COSMOS3. First, ($i$) in Marchesi et al. (2016, Figure 14) we found a good agreement, in the L-COSMOS3 luminosity range, between the fraction of obscured (HR-estimated) sources and the fraction of optically classified non-Type 1 sources. Moreover, ($ii$) we are analyzing the X-ray spectral properties of the 1855 \cha \leg sources having more than 30 net counts in the 0.5-7 keV band (Marchesi et al. in prep.) and we find a general good agreement between the optical and the X-ray classification, e.g., a significant discrepancy between the optical Type 1 and Type 2 intrinsic absorption distributions, the latter having on average three times higher $N_{\rm H}$ values.

\begin{figure*}
  \begin{minipage}[b]{.5\linewidth}
    \centering
  \includegraphics[width=.99\linewidth]{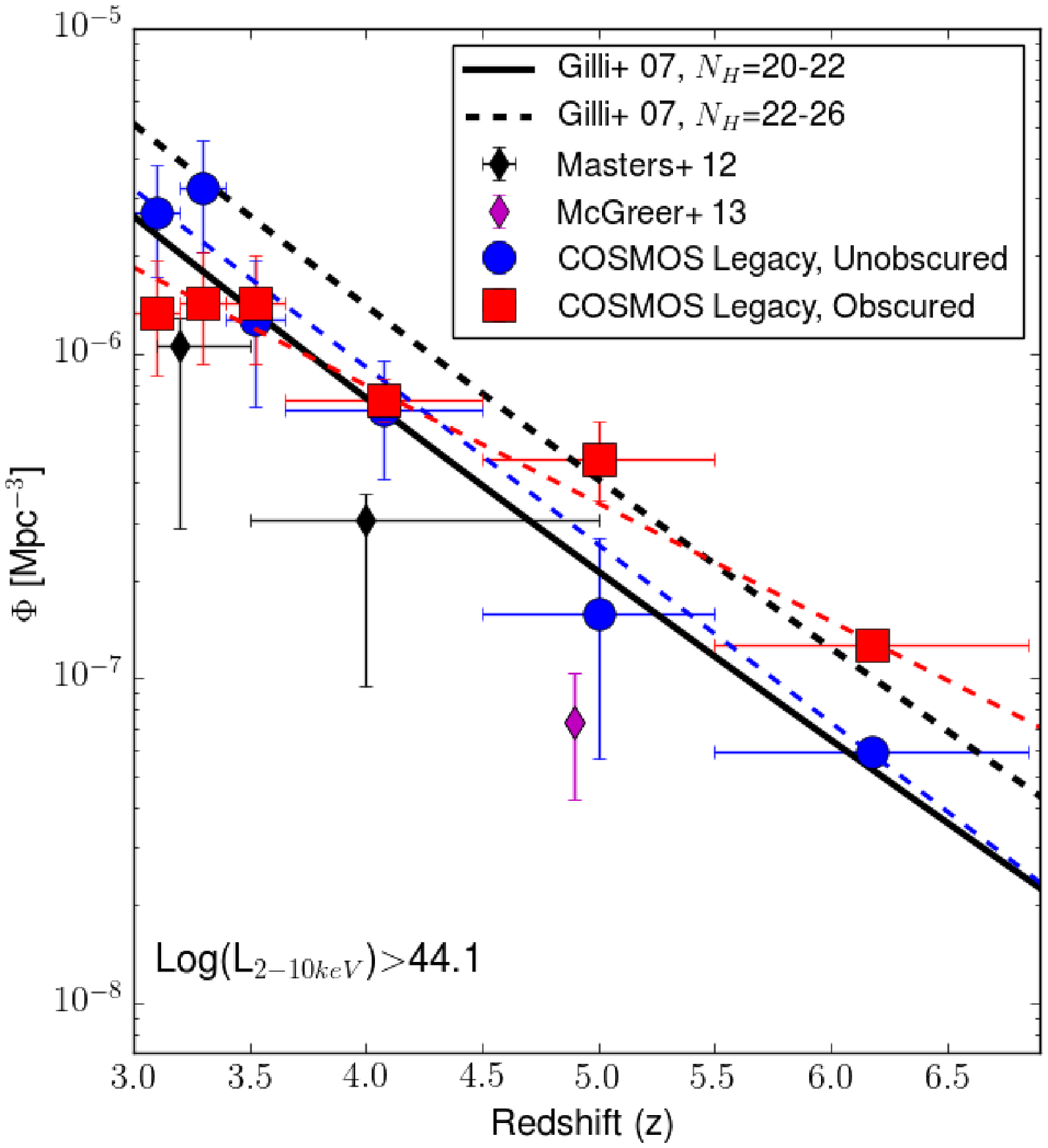}
\end{minipage}%
  \begin{minipage}[b]{.5\linewidth}
    \centering
  \includegraphics[width=.99\linewidth]{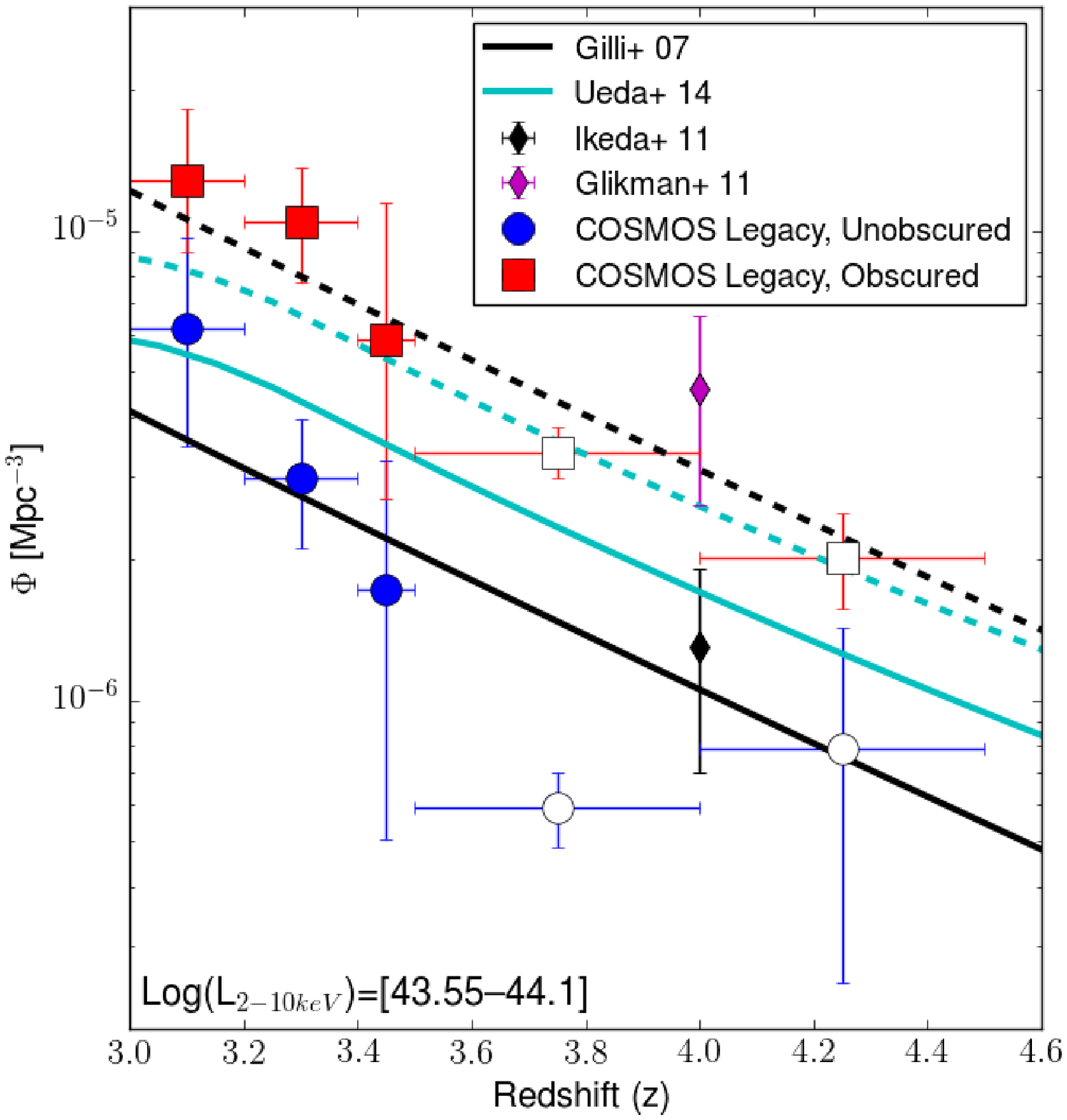}
\end{minipage}
\caption{Space density for sources with $L_{\rm X}$$>$10$^{44.1}$ (left) and 10$^{43.55}$$\leq$$L_{\rm X}$$<$10$^{44.1}$ (right), in the 2-10 keV band. The sample of type 1 AGN is plotted with blue dots, while the sample of type 2 AGN is plotted with red squares. The blue and red dashed lines in the left panel are the best fit to the Type 1 and Type 2 samples, respectively, assuming an equation $Log(\Phi)=a + b \times z $. The model of X-ray population synthesis from Gilli et al. (2007) is also shown, one with $N_H$=[20-22] (black solid line), the other with  $N_H$=[22-26] (black dashed line); the model from Ueda et al. (2014) is shown in cyan, for $N_H$=[20-22] (solid line) and  $N_H$=[22-24] (dashed line). Optical space densities from Masters et al. (2012, black diamonds, left; Type 1 AGN only), McGreer et al. (2013, magenta diamonds, left; Type 1 AGN only), Ikeda et al. (2011, black diamonds, right) and Glikman et al. (2011, magenta diamonds, right)  are also shown for comparison. The open markers in the right panel are the results obtained in those redshift bins where our survey is not complete, and therefore should be treated as lower limits.} 
\label{fig:space_w_type}
\end{figure*}

The space densities for sources with $L_{\rm X}$$>$10$^{44.1}$ (left) and 10$^{43.55}$$\leq$$L_{\rm X}$$<$10$^{44.1}$ (right), in the 2-10 keV band, are shown in Figure \ref{fig:space_w_type}. The sample of type 1 AGN is plotted with blue circles, while the sample of type 2 AGN is plotted with red squares. Our results are also compared with the predictions of the LDDEexp models of Gilli et al. (2007, black lines) and Ueda et al. (2014, cyan line), where the contribution of sources with $N_H$$\leq$10$^{22}$ cm$^{-2}$ (i.e. the unobscured ones) is plotted as a solid line, while the contribution of sources with $N_H$$>$10$^{22}$ cm$^{-2}$ (i.e. the obscured ones) is plotted as a dashed line. At high luminosities (left in Figure \ref{fig:space_w_type}), the unobscured sources ($b$=-0.60$\pm$0.07) are in excellent agreement with the predictions of the model, at any redshift. The trend of decline in obscured sources is instead flatter ($b$=-0.34$\pm$0.04) than the predictions of the model, with the number of obscured sources being smaller than the predictions of the model by a factor $\simeq$2 at $z<$4, while at $z>$4 the data and the model agree. The ratio between obscured and unobscured sources is $\sim$0.4--0.5 in the redshift range $z$=[3-3.4], while it grows to $\sim$1 in the redshift range $z$=[3.4-4] and finally reaches values $\geq$2 at $z\geq$4.5 and above. However, these results need to be verified with a larger sample of spectroscopically verified sources, given that the best-fid SED template classification could be less reliable at these extreme redshifts, where sources are faint in both the optical/IR and the X-ray bands. Moreover, at $z>$5.5, i.e., in the highest redshift bin in our high-luminosity space density, the caveat we described in the previous section (i.e., working only with photo-$z$) must be taken into account.

We also compare our results with those from the optical surveys of Masters et al  (2012, black diamonds, left) and  McGreer et al. (2013, blue diamond, left): there is a good agreement (within 1$\sigma$) between the optical space densities and our unobscured space density, which also have consistent slopes ($b$=-0.68$\pm$0.02 and $b$=-0.60$\pm$0.07 for the unobscured X-ray sources). This result is consistent with our expectations, due to the fact that the optical surveys are limited to Type 1, unobscured sources.

At lower luminosities (10$^{43.55}$$\leq$$L_{\rm X}$$<$10$^{44.1}$, right in Figure \ref{fig:space_w_type}) there are larger uncertainties, but we find that the Type 2 AGN space density is $\sim$2--3 times higher than the Type 1 AGN space density over the whole redshift range $z$=[3-4.5]. Our data are in rough agreement with the predictions of the LDDE models with decline from Gilli et al. (2007, black lines) and Ueda et al. (2014, cyan line), for both unobscured and obscured sources. At $z\sim$4 our unobscured space density fairly agrees with those obtained by Ikeda et al. (2011), obtained using optically selected Type 1 AGN. The result obtained by Glikman et al. (2011) at the same redshift, once again using optically selected Type 1 AGN, lies instead a factor of $\sim$5 above our data. Although our measures at $z\sim$4 are actually lower limits, since the \cha \leg sample is not completed at this redshift and luminosity range, our results challenge those of Glikman et al. (2011).

\begin{table}[H]
\centering
\scalebox{0.82}{
\begin{tabular}{ccc|cc}
\hline
\hline
 &\multicolumn{2}{c}{Type 1} & \multicolumn{2}{c}{Type 2}\\
 & a & b & a & b\\
\hline
Log$L_{\rm X}$$>$44.1 & -3.85$\pm$0.15 & -0.55$\pm$0.03 & -4.65$\pm$0.11 & -0.36$\pm$0.02\\
\hline
43.55$<$Log$L_{\rm X}<$44.1 & -1.46$\pm$1.00 & -1.25$\pm$0.28 & -2.66$\pm$0.33 & -0.74$\pm$0.09\\
\hline
\hline
\end{tabular}}\caption{Parameters of the best fit of the space density for both obscured and unobscured sources, in each range of luminosity, where the fit model is described by the equation $Log(\Phi)=a + b \times z$. In this fit we also take into account the uncertainty on $\Phi$.}\label{tab:fit_w_type}
\end{table}

\section{Comparison with merger models}\label{sec:merger}
Merger-driven models of quasar triggering provide physical framework that fairly well predicts the redshift evolution of the space density of luminous AGN ($L_{bol}>$10$^{46}$ erg s$^{-1}$), with its peak at $z\sim$2-3 and the following decline (e.g., Haiman \& Loeb 1998; Volonteri et al. 2003; Hopkins et al. 2008).

In this section we compare our results with those predicted by the basic quasar activation merger model by Shen (2009). The aim is to use the space density at high redshift to possibly constrain the accretion mechanisms of BH growth and to disentangle between models of BH and galaxy co-evolution. Following Civano et al. (2011) and Allevato et al. (2014), we compare the Shen (2009) merger model with the newest available AGN data at z$>$3, including the ones presented in this work.

The Shen (2009) model was built upon the dark matter halo major merger rate extracted from numerical simulations (Springel et al. 2005; Fakhouri \& Ma 2008), which provides the number of triggering events per unit time, convolved with an assumed AGN light curve, which characterizes the evolution of individual quasars. The light curve is described by an exponentially ascending phase, and a power-law descending phase. The end of the exponential growth is controlled by an AGN feedback self-regulation condition between the peak luminosity and the host dark matter haloes of the type (e.g., Wyithe \& Loeb 2003) $L_{\rm peak} \propto M_{\rm halo}^{5/3}$, valid in the whole range of host halo masses above $M_{\rm halo}>2\times 10^{11}\, M_{\odot}/h$. The parameters of the model were tuned by Shen (2009) to broadly reproduce the full bolometric, obscuration-corrected, AGN luminosity function at 0.1$<$z$<$6, as well as the available large-scale clustering measurements of optical quasars available at the time.

\begin{figure}
\centering
\includegraphics[width=0.5\textwidth]{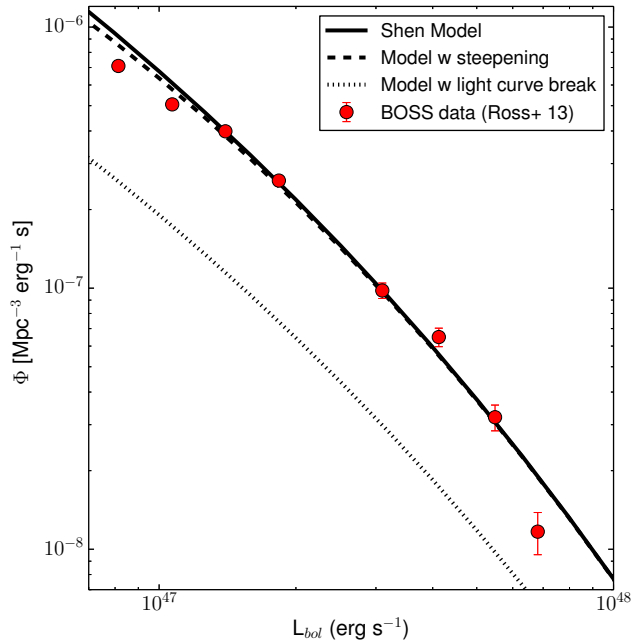}
\caption{SDSS-III BOSS bolometric luminosity function computed in the redshift range $z$=[3.08-3.27] (Ross et al. 2013, red dots). Different models from Shen (2009) are also plotted for comparison: the basic model is plotted as a solid line, the model with a steepening in the Lpeak-$M_{\rm halo}$ relation is plotted with a dashed line and the model with a break in the AGN light curve is plotted with a dotted line. See the text for further details on the modification to the basic model.}\label{fig:lf_boss}
\end{figure}

\begin{figure}
\centering
\includegraphics[width=0.5\textwidth]{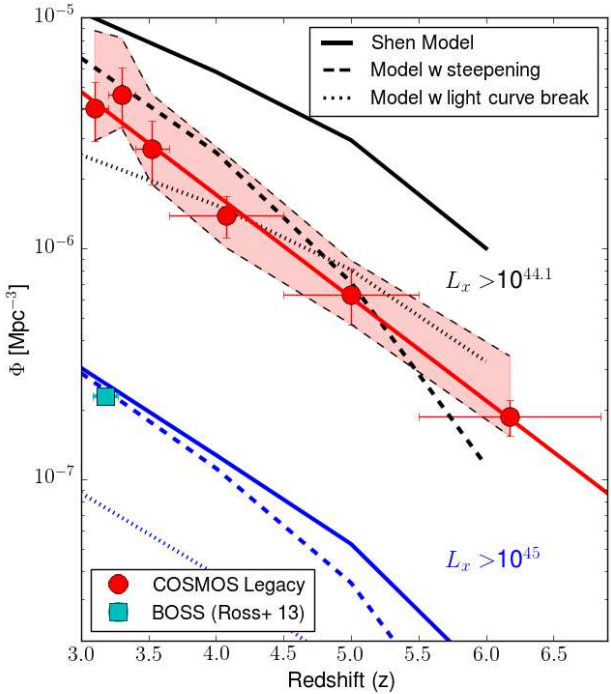}
\caption{\cha \leg space density for sources with $L_{\rm X}$$>$10$^{44.1}$ (red), compared with  different models from Shen (2009, black lines). Space density from BOSS data at $L_{\rm X}>$10$^{45}$ erg s$^{-1}$ (Ross et al. 2013, cyan square) is also plotted, together with different models from Shen (2009, blue lines): the basic model is plotted as a solid line, the model with a steepening in the Lpeak-$M_{\rm halo}$ relation is plotted with a dashed line and the model with a break in the AGN light curve is plotted with a dotted line. See the text for further details on the modification to the basic model.}\label{fig:space_vs_model}
\end{figure}

Figure \ref{fig:lf_boss} shows that the predictions of the reference merger model (black solid line) match well with the high-luminosity part of the optical quasar luminosity function (LF) in the redshift range $z$=[3.08-3.27] by BOSS (Ross et al. 2013). For this comparison, we corrected the model LF by a luminosity-dependent fraction from Ueda et al. (2014) to account only for Type 1 unobscured sources with N$_{H}<$21. The model predictions, however, tend to gradually overestimate the observed space density when moving to fainter luminosities ($L_{bol}<$10$^{47}$ erg s$^{-1}$). This is even more evident when comparing the Shen (2009) model with the number densities of fainter AGN derived in this work (Figure \ref{fig:space_vs_model}). The reference model (black solid line) is higher than the data by a factor of 3 to 10, depending on the redshift. This behavior is not fully unexpected. The Shen (2009) model was calibrated mostly on bright AGN at z$>$3, while the faint AGN data available at the time were poor; it is also worth noticing that such an over-prediction was already observed by Fiore et al. (2012), using the $z>$3 sample from the 4 Ms CDF-S.
 
At fixed redshift, the parameters defining the model seem to be well suited to reproduce the bright end of the AGN luminosity, but tend to fail in matching the most up-to-date number counts from X-ray surveys. There are broadly two ways to improve the match between merger models and data: modify the AGN light curve, or the host halo mass distribution, or a combination of both. 
\begin{enumerate}
\item The black dotted lines in Figures \ref{fig:lf_boss} and \ref{fig:space_vs_model} mark the predictions from a modified Shen (2009) model in which we modified the AGN light curve which characterizes the evolution of individual quasars, described by the combination of an exponential ascending phase and a power-law descending phase. We cut out the post-peak descending phase, with all other parameters held fixed. Cutting out the post-peak descending phase can be physically interpreted as a natural consequence of a powerful quasar feedback, capable of massively clearing out gas from the host galaxy and thus stopping the fueling onto the central black hole (e.g., Granato et al. 2004, Lapi et al. 2006).
This change in the model represents an improvement with respect to the faint-end luminosity function, because a smaller number of low-luminosity AGN is now predicted by the model, though it also tends to cause an under-prediction of the bright-end of the AGN LF. 
\item A second variant to the Shen (2009) model is characterized by a steepening in the $L_{\rm peak}$-$M_{\rm halo}$ relation below $M_{\rm halo}\simeq$10$^{12}$ M$_{\odot}$/h, with $L_{\rm peak} \propto M_{\rm halo}^{5}$ instead of $L_{\rm peak} \propto M_{\rm halo}^{5/3}$, implying that preferentially lower-luminosity quasars are now related to more massive, less numerous host dark matter haloes. In this scenario, less massive black holes within less massive host halos produce a less efficient feedback. This can be caused by gas accretion being less effective in weaker potential wells, which are less effective in retaining gas inside the halo and close to the SMBH (e.g., Kauffmann \& Haenaelt 2000). A second potential cause is a direct correlation between the mass of the black hole and the efficiency of the feedback process (e.g., Granato et al. 2004; Fontanot et al. 2015). In both cases, the final result is a break in the black hole--host galaxy scaling relations.
The outcome of this third model is shown with dashed lines in Figures \ref{fig:lf_boss} and \ref{fig:space_vs_model}. With this model the number densities of very luminous quasars are preserved, while those of lower luminosities ones gradually decrease, in better agreement with the data. Evidence for a break in the black hole-galaxy scaling relations is also now claimed in the local universe (Scott \& Graham 2013), and by independent theoretical models (Cirasuolo et al. 2005, Fontanot et al. 2006, Fontanot et al. 2015). Biases in the local samples of dynamically-measured black holes may however seriously limit our true knowledge of the intrinsic scaling relations between black holes and their host galaxies (e.g., Shankar et al. 2016 and references therein). 
\end{enumerate}

An independent test of the Shen (2009) model will be presented in Allevato et al. (in prep.) making use of the clustering analysis.

\subsection{Alternatives to mergers}
At face value, theoretical merger models predict enough, if not even too many, major mergers to account for all high-redshift AGN of moderate-to-high-luminosity. This does not imply that moderate or minor (e.g., with dwarf galaxies) mergers may not have happened in these systems, given that disk regrowth in gas-rich systems may be a viable possibility at these masses (e.g., Hopkins et al. 2009, Puech et al. 2014, Huertas-Company et al. 2015). Nevertheless our data challenge a pure merger-driven scenario, in agreement with the results of Cisternas et al. (2013), based on galaxy morphology in the local universe.

Mergers may not be the unique driver for the evolution of AGN, especially at lower luminosities. Other ``in-situ'' processes such as disk instabilities and/or clumpy accretion may be effective in channelling flows of gas down towards the very center of the host galaxy, eventually fuelling the black hole (e.g., Bower et al. 2006; Bournaud et al. 2011; Di Matteo et al. 2012). Dedicated studies based on advanced semi-analytic models have shown however that disk instabilities alone may not be enough to account for the full distribution of AGN luminosities (e.g., Menci et al. 2014, Gatti et al. 2015, Gatti et al. 2015 submitted), and in fact direct observations suggest that mergers may be the mechanism driving the most luminous, high-redshift sources (e.g., Treister et al. 2012).

\section{Conclusions}\label{sec:conclusions}
In this work we have selected a sample of 174 $z\geq$3 sources from the \cha \leg survey, the largest sample of $z\geq$3 X-ray selected sources on a contiguous field. 87 of the 174 have a spectroscopic redshift.
We treated the 87 sources with only photometric redshifts as a probability weighted sum, using only the contribution to the PDF at $z\geq$3: the sum of all these contributions is equivalent to have 103.2 sources with $z\geq$3. 66.0 of these sources are from objects with peak of the PDF $z_{peak}\geq$3, while other 37.2 come from a sample of 286 sources with $z_{peak}<$3 but with PDF contribution at $z\geq$3. The final sample is equivalent to 190.2 sources. In the computation of the LogN-LogS and of the space density we used a sample of 179.4 sources, obtained adopting as flux limit the one at which 10\% of the \cha \leg area is covered, to reduce Eddington bias effects.
Here we summarize the main results we obtained.
\begin{enumerate}
\item 85 sources are unobscured Type 1 (49\% of the whole sample, 54 sources with spectral type, the remaining 31 with only photometric type), while the remaining 89 are obscured Type 2 (51\% of the whole sample, 28 sources with spectral type, the remaining 62 with photometric type). The mean (median) $i$-band magnitude is 23.4 (23.4) for Type 1 AGN and 25.3 (25.4) for Type 2 AGN. 
\item Our analysis of the number counts in the observed 0.5-2 keV band shows a decline in the number of sources at $z>$3 and $z>$4 (Figure \ref{fig:logn-logs_1}, left and right panels, respectively). Our results confirm that an exponential decline at redshift $z>$3 is observed in the AGN X-ray number counts, as it is observed in the optical band.
\item For the first time, we were able to put constraints on the number counts at $z$$>$5 (Figure \ref{fig:logn-logs_2}, left) and $z$$>$6 (right). At $z>$6, we measure [0.7--2.2] objects per square degree
\item We computed the rest frame 2-10 keV comoving space density in the high-luminosity range of our survey ($L_{\rm X}$$>$10$^{44.1}$ erg s$^{-1}$, Figure \ref{fig:space1}, left). We observe a decline of a factor $\sim$20 in the space density from $z$=3 to $z$=6.2. Our data are well fitted by a power-law with slope $b$=--0.45$\pm$0.02.
\item In the low luminosity regime ($L_{\rm X}$=[10$^{43.55}$-10$^{44.1}$] erg s$^{-1}$; Figure \ref{fig:space1}, right) the best linear fit to our rest frame 2-10 keV comoving space density ($\Phi$=$a$+$bz$, red solid line) has a slope $b$=--0.82$\pm$0.18, with a steeper decline than the one observed at higher luminosities. 
\item We compared our space density results with the predictions of different phenomenological models. All the phenomenological models have been calibrated at low redshifts and then extrapolated to the high redshift regime we are sampling. At $L_{\rm X}$$>$10$^{44.1}$ erg s$^{-1}$, the FDPL model overpredicts our data by a factor $\sim$2 at 3$<$$z$$<$5, while our data are in good agreement with the predictions of different LDDE models with an exponential decline. Our data are also in good agreement with the results of Vito et al. (2014), which are well fitted by a PDE model. In the low luminosity regime ($L_{\rm X}$=[10$^{43.55}$-10$^{44.1}$] erg s$^{-1}$]), our data seem in slightly better agreement with the LDDE models with exponential decline than with the FDPL model, in the redshift range, $z$=[3-3.4]. 
\item We investigated the 2-10 keV space density for optically classified Type 1 (or unobscured) and Type 2 (or obscured) AGN (Figure \ref{fig:space_w_type}). We found that at $L_{\rm X}$$>$10$^{44.1}$ erg s$^{-1}$ obscured sources have a slope significantly flatter ($b$=--0.34$\pm$0.04) than unobscured sources ($b$=--0.60$\pm$0.07). The ratio between obscured and unobscured sources is $\leq$1 in the redshift range $z$=[3-4], while it grows to $\simeq$2 at $z$=5.
\item We compared our data with the quasar activation merger models of Shen (2009),  caliibrated mostly on luminous Type 1 AGN at z$>$3. We found that the model significantly overpredicts, by a factor of 3--10, with respect to our space density data. To find a closer agreement between data and model, we imposed that most of z$>$3 AGN are preferentially hosted in more massive haloes. This change in the model predicts a specific clustering pattern that we are testing and we will discuss in a future work (Allevato et al. in prep.).
\end{enumerate}

We point out that in this work we did not analyze the basic X-ray properties of our sample (e.g., the hardness ratio). However, we are going to perform a detailed analysis of the X-ray spectral properties (i.e., spectral slope, obscuration, evidence of iron $K \alpha$ emission lines) of the $\simeq$2000 sources \cha \leg sources with more than 30 net counts in the 0.5-7 keV (Marchesi et al. in prep.). In this same work, we will discuss in detail the X-ray properties of the L-COSMOS3 sample.

We briefly summarize several other projects, already submitted or in preparation, based on the L-COSMOS3 dataset and on the results presented in this work. 

\begin{enumerate}
\item A spectroscopic follow-up of two of the four candidate $z>$6 sources in L-COSMOS3 will be performed in early 2016 (P.I.: F. Civano) using \textit{Keck}-LRIS. If one of these redshifts would be confirmed, this would be the first spectroscopically confirmed X-ray selected AGN at $z>$6. 
\item A subsample of 10 bright sources from LCOSMOS-3 at $z\sim$3.3 has already been observed with \textit{Keck} MOSFIRE, allowing to estimate the BH mass and put better constraints on the accretion properties of SMBH in early universe (Trakhtenbrot et al. 2015b submitted). The AGN in this subsample are powered by SMBHs with M$_{BH}\sim$ 6 $\times$ 10$^8$ M$_\odot$ and L/L$_{Edd}\sim$0.1-0.5. Fainter sources may be powered by lower-mass and/or accretion rate SMBHs. One of these 10 sources, CID\_947 ($z$=3.328), showed an extremely massive accreting BH, with $M_{BH}$$\simeq$0.1 $M_{galaxy}$, therefore suggesting a much faster BH mass accretion than that of the host galaxy (Trakhtenbrot et al. 2015a). 
\item A work on the clustering properties of the \cha \leg $z>$3 sample is being performed (Allevato et al. in prep.), to study properties such as AGN radiative efficiency and Eddington ratio, and the black hole duty cycle (Shankar et al. 2010a,b; Allevato et al 2014). 
\item The L-COSMOS3 space density is being used to study the AGN UV emissivity and to estimate the contribution of AGN to the reionization of the Universe at $z>$6  (Ricci et al. submitted).
\end{enumerate}

Finally, we point out that only future facilities like the X-ray Surveyor will be able to collect sizable samples of low luminosities ($<$10$^{43}$ erg s$^{-1}$) AGN at $z>$5 (Civano et al. 2015).

\section{Acknowledgments}
We thank the anonymous referee for the comments and suggestions that helped in improving the paper. This research has made use of data obtained from the \cha Data Archive and software provided by the \cha X-ray Center (CXC) in the CIAO application package.

This work was supported in part by NASA Chandra grant number GO3-14150C and also GO3-14150B  (F.C., S.M., V.A., M.E.); PRIN-INAF 2014 "Windy Black Holes combing galaxy evolution" (A.C., M.B., G.L. and C.V.); the FP7 Career Integration Grant ``eEASy'': ``Supermassive black holes through cosmic time: from current surveys to eROSITA-Euclid Synergies" (CIG 321913; M.B. and G.L.); UNAM-DGAPA Grant PAPIIT IN104216 and  CONACyT Grant Cient\'ifica B\'asica \#179662 (T.M.); NASA award NNX15AE61G (R.G.);  the Swiss National Science Foundation Grant PP00P2\_138979/1 (K.S.); the Center of Excellence in Astrophysics and Associated Technologies (PFB 06), by the FONDECYT regular grant 1120061 and by the CONICYT Anillo project ACT1101 (E.T.). B.T. is a Zwicky Fellow.

\end{document}